\documentclass[aps,prb,twocolumn,showpacs,amsmath,amssymb,superscriptaddress,floatfix]{revtex4-2}

\usepackage[english]{babel}
\usepackage{blindtext}
\usepackage{latexsym}
\usepackage{amssymb}
\usepackage{physics}
\usepackage{amsmath}
\usepackage{bm}
\usepackage{mathtools}
\usepackage{amsfonts}
\usepackage{relsize}
\usepackage{xcolor}
\usepackage{verbatim}
\usepackage{bbold}
\usepackage{slashed}
\usepackage{appendix} 
\usepackage{graphicx}
\usepackage[normalem]{ulem}
\usepackage[colorlinks = true,
            linkcolor = blue,
            urlcolor  = blue,
            citecolor = blue,
            anchorcolor = blue]{hyperref}
\usepackage{graphicx} % Include figure files
\usepackage{dcolumn} % Align table columns on decimal point
\newcommand{\approptoinn}[2]{\mathrel{\vcenter{
  \offinterlineskip\halign{\hfil$##$\cr
    #1\propto\cr\noalign{\kern2pt}#1\sim\cr\noalign{\kern-2pt}}}}}

\newcommand*{\colorboxed}{}
\def\colorboxed#1#{%
  \colorboxedAux{#1}%
}
\newcommand*{\colorboxedAux}[3]{%
  \begingroup
    \colorlet{cb@saved}{.}%
    \color#1{#2}%
    \boxed{%
      \color{cb@saved}%
      #3%
    }%
  \endgroup
}

\newcommand{\up}{\uparrow}
\newcommand{\down}{\downarrow}
\newcommand{\bs}[1]{\mathbf{#1}}

\newcommand{\bq}{\bs{q}}
\newcommand{\bQ}{\bs{Q}}
\newcommand{\bzero}{\bs{0}}
\newcommand{\bk}{\bs{k}}

\newcommand{\chit}{\widetilde{\chi}}

\definecolor{ao(english)}{rgb}{0.0, 0.5, 0.0}
\definecolor{amaranth}{rgb}{0.9, 0.17, 0.31}
\definecolor{green(html/cssgreen)}{rgb}{0.0, 0.5, 0.0}

\newcommand\greensout{\bgroup\markoverwith{\textcolor{green(html/cssgreen)}{\rule[0.5ex]{2pt}{1.0pt}}}\ULon}

\begin{document}
\title{Spiral to stripe transition in the two-dimensional Hubbard model}
\author{Robin Scholle}
\affiliation{Max Planck Institute for Solid State Research, D-70569 Stuttgart, Germany}
\author{Walter Metzner}
\affiliation{Max Planck Institute for Solid State Research, D-70569 Stuttgart, Germany}
\author{Demetrio Vilardi}
\affiliation{Max Planck Institute for Solid State Research, D-70569 Stuttgart, Germany}
\author{Pietro M.~Bonetti}
\affiliation{Max Planck Institute for Solid State Research, D-70569 Stuttgart, Germany}
\affiliation{Department of Physics, Harvard University, Cambridge, Massachusetts 02138, USA}
\date{\today}
\begin{abstract}
We obtain an almost complete understanding of the mean-field phase diagram of the two-dimensional Hubbard model on a square lattice with a sizable next-nearest neighbor hopping and a moderate interaction strength. In particular, we clarify the nature of the transition region between the spiral and the stripe phase. Complementing previous [Phys.\ Rev.\ B {\bf 108}, 035139 (2023)] real-space Hartree-Fock calculations on large finite lattices, we solve the mean-field equations for coplanar unidirectional magnetic order directly in the thermodynamic limit, and we determine the nature of the magnetic states right below the mean-field critical temperature $T^*$ by a Landau free energy analysis. While the magnetic order for filling factors $n \geq 1$ is always of N\'eel type, for $n \leq 1$ the following sequence of magnetic states is found as a function of increasing hole-doping: N\'eel, planar circular spiral, multi-spiral, and collinear spin-charge stripe states. Multi-spiral states are superpositions of several spirals with distinct wave vectors, and lead to concomitant charge order.
We finally point out that nematic and charge orders inherited from the magnetic order can survive even in the presence of fluctuations, and we present a corresponding qualitative phase diagram.
\end{abstract}
\maketitle

%%%%%%%%%%%%%%%%%%%%%%%%%%%%%%%%%%%%%%%%%%%%%%%%%%%%%%%%%%%%%%%%%%%%%%%%%%%%%%%%%%%%

\section{Introduction}

The two-dimensional Hubbard model on a square lattice plays a key role in the field of strongly correlated electron systems as a prototype model for competing and intertwined ordering tendencies. It captures the most salient features of electrons in the copper oxide planes of high-$T_c$ cuprates, namely antiferromagnetism and $d$-wave superconductivity \cite{Scalapino2012}.
Thanks to remarkable advances in the development of computational techniques, fragments of the phase diagram of this important model have been established \cite{Arovas2022, Qin2022}, but many regions in the large parameter space spanned by hopping amplitudes, interaction strength, electron filling, and temperature remain {\em terra incognita}.

In the most interesting (broad) density range around half-filling, there is a competition and possible coexistence of magnetic order, charge order, and superconductivity.
While plausible candidates for superconducting states are easily classified, there is an overwhelming zoo of possible magnetic states.
At half-filling, the ground state of the Hubbard model is a simple N\'eel antiferromagnet.
Away from half-filling, most calculations indicate either planar circular spin spirals \cite{Shraiman1989, Shraiman1992, Chubukov1992, Chubukov1995, Dombre1990, Fresard1991, Kotov2004, Igoshev2010, Igoshev2015, Yamase2016, Eberlein2016, Mitscherling2018, Bonetti2020a} or spin-charge stripes with collinear spin order and concomitant charge order \cite{Schulz1989, Zaanen1989, Machida1989, Poilblanc1989, Schulz1990, Kato1990, Seibold1998, Fleck2000, Fleck2001, Raczkowski2010, Timirgazin2012, Peters2014, Matsuyama2022, Zheng2017, Qin2020} as energetically favorable magnetic alternatives to the N\'eel state.

A large variety of magnetic phases in the two-dimensional Hubbard model emerges already in a conventional static mean-field approximation, that is, Hartree-Fock theory.
While the regime of ordered states in the phase diagram is usually overestimated by mean-field theory, qualitative insights may serve as a guide for more sophisticated calculations, in particular, to interprete data from numerical simulations on finite lattices.
Numerous Hartree-Fock studies of the two-dimensional Hubbard model have already been published. In many of them the magnetic order was restriced to certain patterns, such as ferromagnetic and N\'eel order \cite{Lin1987, Hofstetter1998, Langmann2007} or, more generally, to spiral order with arbitrary wave vectors \cite{Igoshev2010, Igoshev2015}. Allowing for collinear magnetic order with arbitrary wave vectors or even for completely arbitrary spin configurations, spin-charge stripes have been discovered \cite{Schulz1989, Zaanen1989, Machida1989, Poilblanc1989, Schulz1990, Kato1990}.

Mean-field theory yields magnetic order also at finite temperatures, below a transition temperature $T^*$, violating thus the Mermin-Wagner theorem \cite{Mermin1966}. However, magnetically ordered states at finite temperature become meaningful in theories of fluctuating magnetic order, where the electron is fractionalized into a fermion with a magnetically ordered pseudospin, and a fluctuating SU(2) rotation matrix which restores the SU(2) spin symmetry \cite{Scheurer2018, Sachdev2019, Bonetti2022gauge}.

Recently, we have performed a comprehensive and unbiased mean-field analysis of magnetic and charge order in the Hubbard model with a moderate interaction strength on a square lattice, at both zero and finite temperatures \cite{Scholle2023}. Completely unrestricted real-space Hartree-Fock calculations on large finite lattices were combined with a stability analysis of mean-field solutions restricted to N\'eel and spiral order in the thermodynamic limit. It turned out that in most parts of the phase diagram only three classes of magnetic states with a relatively simple structure are stabilized in the thermodynamic limit: N\'eel, circular spiral, and collinear stripe states. The stripes are usually unidirectional, but can also be bidirectional at very large hole doping in presence of a sizable next-nearest neighbor hopping. In spite of rather large lattices (up to $48 \times 48$) used in the real-space calculations, the analysis of the stripe states was still hampered by finite size effects, and the nature of the transition from the spiral to the stripe phase remained open.

In this paper we complete the mean-field analysis of our previous work \cite{Scholle2023} by performing several complementary calculations. First, we solve the mean-field equations for ground states with generic coplanar unidirectional order directly in the thermodynamic limit. This includes the N\'eel, spiral, and unidirectional stripe states found in the real-space calculations \cite{Scholle2023} as special cases. Second, we determine the magnetic ordering pattern right below $T^*$ from a Landau free energy analysis. Third, we clarify the nature of the instability of the spiral state upon increasing doping by analyzing its spin susceptibility, again directly in the thermodynamic limit. We find that the transition from the spiral to the stripe phase leads through a rather complex intermediate phase with a superposition of multiple spiral components with three or four distinct wave vectors. Finally, we present a qualitative discussion of fluctuation effects. Order parameter fluctuations restore the SU(2) spin symmetry at least at finite temperature, but nematic and charge orders found in the mean-field states may survive.

The remainder of our paper is structured as follows. In Sec.~\ref{sec: ModelMethod} we describe our three complementary methods used to compute the mean-field phase diagram and to clarify the nature of the various magnetic states. In Sec.~\ref{sec: results} we present the corresponding results. In the Conclusion in Sec.~\ref{sec: Conclusion} we summarize and present a qualitative discussion of fluctuation effects.

%%%%%%%%%%%%%%%%%%%%%%%%%%%%%%%%%%%%%%%%%%%%%%%%%%%%%%%%%%%%%%%%%%%%%%%%%%%%%%%%%%%%

\section{Model and method}\label{sec: ModelMethod}
The Hubbard Hamiltonian for spin-$\frac{1}{2}$ fermions with intersite hopping amplitudes $t_{jj'}$ and a local repulsive interaction $U>0$ reads \cite{Arovas2022, Qin2022}
\begin{equation} \label{eq: HubbardHamiltonian}
 H = 
 \sum_{j,j',\sigma} t_{jj'} c^\dagger_{j\sigma} c^{\phantom\dagger}_{j'\sigma} +
 U \sum_j n_{j\uparrow} n_{j\downarrow} \, ,
\end{equation}
where $c_{j\sigma}$ ($c^\dagger_{j\sigma}$) annihilates (creates) an electron on lattice site $j$ with spin orientation $\sigma \in \{\uparrow,\downarrow\}$, and $n_{j\sigma} = c^\dagger_{j\sigma} c^{\phantom\dagger}_{j\sigma}$. The hopping matrix $t_{jj'}$ depends only on the distance between the sites $j$ and $j'$. We choose $t_{jj'} = -t$ if $j$ and $j'$ are nearest neighbor sites, $t_{jj'} = -t'$ if $j$ and $j'$ are next-to-nearest neighbors, and $t_{jj'} = 0$ otherwise. We use the nearest neighbor hopping amplitude $t$ as our energy unit.

In mean-field theory, the interaction term in~\eqref{eq: HubbardHamiltonian} can be decoupled as~\cite{Zaanen1989,Scholle2023}
\begin{equation}\label{eq: MF decoupling}
 \begin{split}
 U \sum_j n_{j\uparrow} n_{j\downarrow} \simeq \sum_{j}\sum_{a=0}^3 s_a \left
 [\Delta^a_j \, c^\dagger_j \sigma^a c_j - \frac{(\Delta_j^a)^2}{U} \right] \, ,
 \end{split}
\end{equation}
where $c_j$ is the two-component spinor composed of $c_{j\up}$ and $c_{j\down}$, while $\sigma^0$ is the two-dimensional identity matrix and $\sigma^1,\sigma^2,\sigma^3$ are the Pauli matrices. The sign $s_a$ is plus one if $a=0$, and minus one otherwise.
The parameters $\Delta^a_j$ are related to charge and spin expectation values as
\begin{equation}
 \Delta^a_j = \frac{1}{2} U \langle c_j^\dagger \sigma^a c_j \rangle \, .
\end{equation}
The mean-field decoupling in Eq.~\eqref{eq: MF decoupling} captures both the Hartree ($a=0,3$) and the Fock ($a=1,2$) terms.

Previous unbiased and unrestricted real-space mean-field calculations on the Hubbard model~\cite{Scholle2023} revealed that, except for very low electron densities, the solutions of the mean-field equations always converge to coplanar unidirectional phases. Thus, in this paper we focus on mean-field states characterized by one or more wave vectors of the form $\bQ = (\pi-\delta,\pi)$ or symmetry related (we call this property unidirectionality), and where all the spins lie in a common plane (coplanarity). This includes collinear spin states as special cases (with infinitly many common planes), and in particular the N\'eel state as the collinear state with $\bQ = (\pi,\pi)$.

We analyze the different phases that one can obtain within mean-field theory in the Hubbard model by employing three distinct but mutually consistent techniques.

A) To find the magnetic ground state, we employ a Hartree-Fock ansatz that allows for a generic coplanar unidirectional state with an arbitrary integer periodicity $P$ in $x$-direction (and antiferromagnetic order in $y$-direction).
The contributing wave vectors have the form $\bQ = (2\pi n/P,\pi)$, with integer numbers $n \geq 0$. The mean-field equations are solved directly in the thermodynamic limit.
The periodicity $P$ determines the size of the real-space unit cell one has to deal with in the solution of the mean-field equations. We could reach unit cells as big as 220 sites along the $x$-axis, so that even incommensurate states without any translation symmetry (that is, $P = \infty$) are approximated very well.

B) Close to the critical temperature $T^*$ at which mean-field magnetic order sets in, it is technically hard to obtain converged solutions of the mean-field equations. Therefore, to determine the pattern of the magnetic order setting in right below $T=T^*$, we employ a Landau theory for mixed spin-charge order parameters, and microscopically compute its coefficients from the paramagnetic state at $T=T^*$. Note that in the limit of a vanishing order parameter, Landau theory and Hartree-Fock theory yield the same type of order.

C) It has been previously observed \cite{Shraiman1989, Shraiman1992, Chubukov1992, Chubukov1995, Dombre1990, Fresard1991, Kotov2004, Igoshev2010, Igoshev2015, Yamase2016, Eberlein2016, Mitscherling2018, Bonetti2020a, Scholle2023} that a circular spiral magnetic state is favored for small hole doping (that is, slightly below half-filling) in presence of a finite $t'<0$. We study the instabilities of the spiral state to other magnetic orders at larger hole dopings by computing the spin and charge susceptibilities in such a state within random phase approximation (RPA). This enables us to determine not only when the spiral state becomes unstable, but also the nature of the magnetic order emerging beyond the instability line.
Note that the RPA is the unique conserving approximation for susceptibilities which is consistent with mean-field theory for the free energy, order parameters, and single-particle properties \cite{Baym1961}.

In the following, we provide a detailed description of the three methods mentioned above. 

%%%%%%%%%%%%%%%%%%%%%%%%%%%%%%%%%%%%%%%%%%%%%%%%%%%%%%%%%%%%%%%%%%%%%%%%%%%%%%%%%%%%%%%%

\subsection{Mean-field theory for a generic coplanar unidirectional magnetic state}
\label{sec: MFT}

We derive mean-field equations in momentum space which describe generic magnetic states characterized by the following three properties.

\textit{Coplanarity}. The onsite magnetization should lie in a specific plane, which we choose, without loss of generality, to be the $xy$-plane.

\textit{Unidirectionality}. Spins sitting on neighboring sites along the, say, $y$-direction are antiparallel and the charge densities are equal, while along the $x$-direction the magnetization amplitude and orientation, as well as the charge density, can be arbitrarily modulated.

\textit{Commensurability}. Spin and charge orders display a periodicity with respect to translations along the $x$-axis with a finite integer period (denoted by $P$) in units of the lattice spacing.
This criterion implies a restriction to states with ordering wave vectors commensurate with the lattice. Incommensurate states can be approximated to any desired accuracy by choosing a sufficiently large value for $P$.

In a coplanar, unidirectional, commensurate state with periodicity $P$, the magnetization and charge density profiles can be expressed as follows
\begin{subequations} \label{eq: spin and charge profiles}
\begin{align}
 &\Vec{S}_j = \sum_{n\text{ odd}} \left(M^x_n \hat{e}_x + M^y_n \hat{e}_y \right)
 e^{i n \bQ_P \cdot \mathbf{R}_j} \, , \\
 &\rho_j = \sum_{n\text{ even}} \, \varrho_n\, e^{i n \bQ_P \cdot \mathbf{R}_j} \, ,
\end{align}
\end{subequations}
where $\mathbf{R}_j$ are the coordinates of lattice site $j$, and $\bQ_P = (2\pi/P,\pi)$.
The first sum is running only over odd integers $n$ because the spin order is antiferromagnetic in $y$-direction, while the second sum is restricted to even integers since $\rho_j$ is translation invariant in $y$-direction.
We define $P_S$ as the smallest positive integer satisfying $P_S \bQ_P = (0,0)$ modulo reciprocal lattice vectors. If $P$ is even, one has $P_S = P$, and $P_S = 2P$ if $P$ is odd.
Because $n\bQ_P$ is equivalent to $(n + mP_S)\bQ_P$ with $m,n \in \mathbb{Z}$, the summations in Eq.~\eqref{eq: spin and charge profiles} run only over a finite number ($P_S/2$) of terms. Moreover, since the spin and charge densities on the left hand side of Eqs.~\eqref{eq: spin and charge profiles} are real, the coefficients $M^x_n$, $M^y_n$, and $\varrho_n$ must obey
\begin{subequations}
\begin{align}
 M_n^x & = (M_{-n}^x)^* = (M_{P_S-n}^x)^* \, , \\
 M_n^y & = (M_{-n}^y)^* = (M_{P_S-n}^y)^* \, , \\
 \varrho_n & = (\varrho_{-n})^* = (\varrho_{P_S-n})^* \, .
\end{align}
\end{subequations}

\begin{figure*}[tb]
\centering
\includegraphics[width=1.0\textwidth]{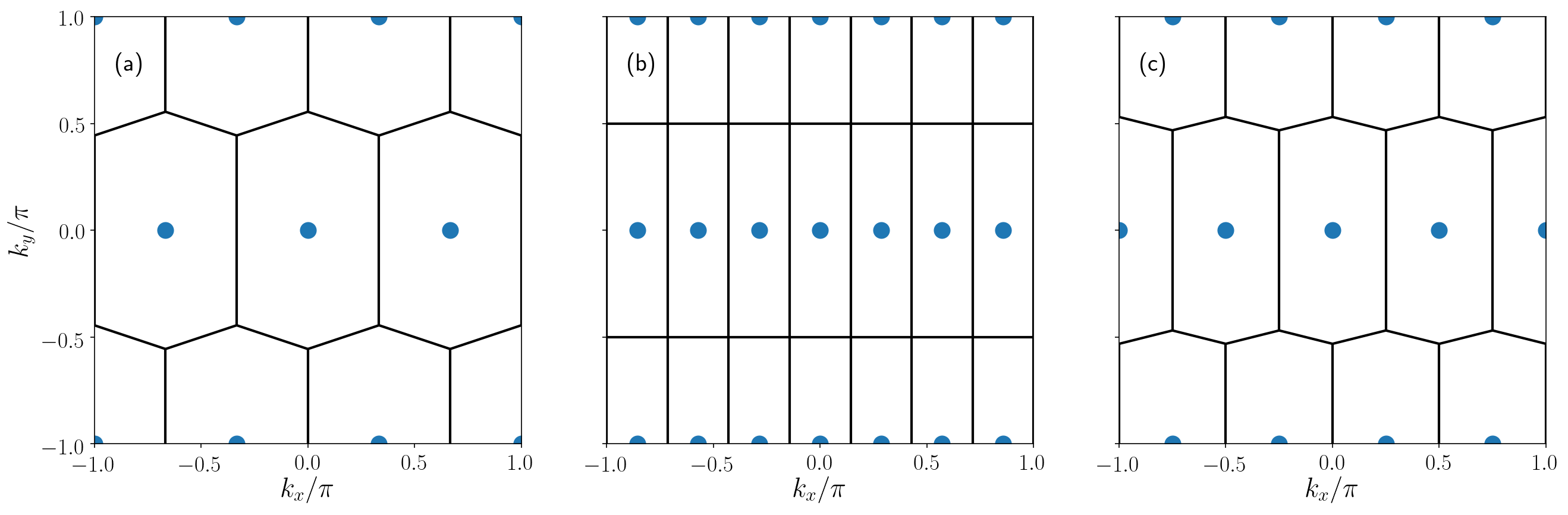}
 \caption{Construction of the reduced Brillouin zone for $P=6$ (a), $P=7$ (b), and $P=8$ (c). The blue dots represent the momenta $n\bQ_P$ with $n=0,...,P_S-1$ modulo vectors of the original reciprocal lattice. The black lines separate distinct reduced Brillouin zones. The original Brillouin zone is divided into $P_S$ equivalent reduced Brillouin zones.}
\label{fig: reduced BZ}
\end{figure*}

Spin and charge orderings break the translational symmetry of the original lattice, resulting in an enlarged unit cell containing $P_S$ inequivalent sites with distinct expectation values. Similarly, in momentum space, the size of the original Brillouin zone is reduced by a factor $1/P_S$. The new reciprocal lattice can be constructed by adding all vectors of the form $n\bQ_P$ with $n=0,...,P_S-1$ to the original reciprocal lattice vectors (see Fig.~\ref{fig: reduced BZ}). The reduced Brillouin zone can then be defined as the set of all points that are closer to a given vector of the new reciprocal lattice than to any other (Wigner-Seitz construction).
The reduced Brillouin zones for $P=6,7,8$ are plotted in Fig.~\ref{fig: reduced BZ}.

Inserting Eq.~\eqref{eq: spin and charge profiles} into the Hamiltonian~\eqref{eq: HubbardHamiltonian} with the mean-field decoupling~\eqref{eq: MF decoupling}, and Fourier transforming, we obtain the quadratic Hamiltonian
\begin{equation}\label{eq. H MF mom space}
\begin{split}
 H_\mathrm{MF} =& \int_\bk \epsilon_\bk\, c^\dagger_\bk c_\bk +
 \sum_{a=0}^2 \sum_{n=0}^{P_S-1}
 s_a \Delta_n^a \, c^\dagger_\bk \sigma^a c_{\bk+n\bQ_P} \\
 & - \sum_{a=0}^2 \sum_{n=0}^{P_S-1} s_a \frac{|\Delta_n^a|^2}{U} \, ,
    \end{split}
\end{equation}
where $\epsilon_\bk = -2t (\cos k_x + \cos k_y ) - 4 t' \cos k_x \cos k_y $ is the Fourier transform of the hopping parameters in Eq.~\eqref{eq: HubbardHamiltonian}. We have also defined
\begin{subequations}
\begin{align}
 \Delta_n^0 &= \left\{ \begin{array}{lll}
  \frac{1}{2} U \varrho_n & \mbox{for} & \mbox{$n$ even} \\
  0             & \mbox{for} & \mbox{$n$ odd} \end{array} \right. \, , \\
  \Delta_n^{1,2} &= \left\{ \begin{array}{lll}
  0           & \mbox{for} & \mbox{$n$ even} \\
  U M_n^{x,y} & \mbox{for} & \mbox{$n$ odd} \end{array} \right. \, .
\end{align}
\end{subequations}
$\int_\bk$ is a shorthand for the integral
$\int_{\bk\in\mathrm{BZ}} \frac{d^2\bk}{(2\pi)^2}$ over the original Brillouin zone (BZ).

Introducing a "Nambu spinor" with $P_S$ components
\begin{equation}
 \Psi_{\bk,\sigma} = \left(
 \begin{array}{c}
 c_{\bk,\sigma} \\
 c_{\bk+\bQ_P,\Bar{\sigma}} \\
 c_{\bk+2\bQ_P,\sigma} \\
 \vdots \\
 c_{\bk+(P_S-2)\bQ_P,\sigma} \\
 c_{\bk+(P_S-1)\bQ_P,\bar{\sigma}}
 \end{array} \right) ,
\end{equation}
with the convention $\bar\up = \, \down$ and $\bar\down = \, \up$, one can cast the Hamiltonian~\eqref{eq. H MF mom space} in the form
\begin{equation} \label{eq: HMF explicit}
 H_\mathrm{MF} =
 \sum_\sigma \int_\bk' \Psi^\dagger_{\bk,\sigma} \mathcal{H}^{(P)}_{\bk,\sigma}
 \Psi_{\bk,\sigma} ,
\end{equation}
where we have dropped the constant term in Eq.~\eqref{eq. H MF mom space} and defined
$\int_\bk'=\int_{\bk\in\mathrm{BZ'}}\frac{d^2\bk}{(2\pi)^2}$,
with $\mathrm{BZ'}$ the reduced Brillouin zone.
The matrix $\mathcal{H}^{(P)}_{\bk,\sigma}$ has the form
\begin{equation} \label{eq: H_MF components}
 \left[\mathcal{H}^{(P)}_{\bk,\sigma}\right]_{\ell\ell'} =
 \left\{ \begin{array}{lll}
 \epsilon_{\bk+\ell\bQ_P} & \text{if} & \ell=\ell' \\
 \Delta^{p(\ell)}_{\sigma,n} & \text{if} &
 \ell'= (\ell+n)\,\mathrm{mod}\,P_S, \, \text{$n$ odd} \\
 \Delta_n^0 & \text{if} &
 \ell'= (\ell+n) \,\mathrm{mod} \, P_S, \, \text{$n$ even}
 \end{array} \right. ,
\end{equation}
where we have defined $p(\ell) = +$ if $\ell$ is even and $p(\ell) = -$ if $\ell$ is odd, and
\begin{equation}
\begin{split}
 \Delta^\pm_{\up,n} &= - (\Delta_n^1 \pm i \Delta_n^2) \, , \\
 \Delta^\pm_{\down,n} &= - (\Delta_n^1 \mp i \Delta_n^2) \, .
\end{split}
\end{equation}
Since $\Delta_0^0$ only shifts the chemical potential $\mu$, in the following we redefine $\mu$ as $\mu - \Delta_0^0$ and set $\Delta_0^0 = 0$.
In Appendix~\ref{app: explicit Hamiltonian} we report the explicit form of the matrix for the cases $P=3$ and $6$.

The parameters $\Delta_n^a$ are self-consistently determined as
\begin{eqnarray} \label{eq: MF gap equations}
 \Delta_n^a &=&
 \frac{1}{2} U \int_\bk \langle c^\dagger_\bk\sigma^a c_{\bk+n \bQ_P} \rangle
 \nonumber \\
 &=& \frac{1}{2} U \sum_\sigma\int_\bk^{\prime}
 \langle \Psi^\dagger_{\bk,\sigma} \Gamma_{n,\sigma}^a \Psi_{\bk,\sigma}\rangle \, ,
\end{eqnarray}
where the matrices $\Gamma^{a,n}_\sigma$ have been defined as
\begin{subequations}
\begin{align}
 & \left[ \Gamma_{n,\sigma}^0 \right]_{\ell\ell'} =
 \begin{cases}
 1 \; \mbox{if $\ell' = (\ell+n)$mod$P_S$, $n$ even} \\
 0 \; \mbox{otherwise}
 \end{cases} \\
 & \left[ \Gamma_{n,\sigma}^1 \right]_{\ell\ell'} =
 \begin{cases}
 1 \; \mbox{if $\ell' = (\ell+n)$mod$P_S$, $n$ odd} \\
 0 \; \mbox{otherwise}
 \end{cases} \\
 & \left[\Gamma_{n,\sigma}^2 \right]_{\ell\ell'} =
 \begin{cases}
 i (-1)^{\sigma + p(\ell)} \; \mbox{if $\ell' = (\ell+n)$mod$P_S$, $n$ odd} \\
 0 \; \mbox{otherwise}
 \end{cases}
\end{align}
\end{subequations}
with $(-1)^\up = +1$ and $(-1)^\down = -1$.

Since $\mathcal{H}_{\bk,\up}^{(P)}$ and $\mathcal{H}_{\bk,\down}^{(P)}$ are related to each other by an inversion of the sign of $\Delta_n^2$, the expectation values on the right hand side of Eq.~\eqref{eq: MF gap equations} take the same value for each of the two spin projections. For this reason, one can simplify Eq.~\eqref{eq: MF gap equations} to
\begin{equation}\label{eq: self-consistent MF equation}
 \Delta_n^a =
 U \int_\bk^{\prime} \langle \Psi^\dagger_{\bk} \Gamma_n^a \Psi_{\bk}\rangle\, ,
\end{equation}
where $\Psi_\bk = \Psi_{\bk,\up}$ and $\Gamma_n^a = \Gamma_{n,\up}^a$. In other words, we can solve the mean-field equations using only the matrix $\mathcal{H}^{(P)}_{\bk,\up}$. The right hand side of Eq.~\eqref{eq: self-consistent MF equation} is computed from Eq.~\eqref{eq: H_MF components} making an initial random assumption on the mean-field parameters $\Delta_n^a$, which are then updated using again Eq.~\eqref{eq: self-consistent MF equation}. The procedure is repeated until convergence is reached.

To find the energetically best state, we converge Eq.~\eqref{eq: self-consistent MF equation} for different values of $P$ and retain the state with the lowest mean-field free-energy. In practice, we discretize the original Brillouin zone BZ with $N_k^2$ equally spaced points and, for a fixed $N_k$, we only allow values of $P$ that are divisors of $N_k$. For every fixed set of parameters, we have offered the system over 90 integer values of $P$ ranging from 2 to 220, each of them with a suitably adjusted $N_k$ such that $N_k/P\in\mathbb{N}$, with $N_k$ ranging from 116 to 220.

In the following, we discuss how several important familiar phases are captured as special cases within our general formalism.

%%%%%%%%%%%%%%%%%%%%%%%%%%%%%%%%%%%%%%%%%%%%%%%%%%%%%%%%%%%%%%%%%%%%%%%%%%%%%%%%%%

\subsubsection{N\'eel order}

In the case of N\'eel order, one has
\begin{subequations}
\begin{align}
 \vec{S}_j &=
 M (-1)^{j} [\cos\varphi \, \hat{e}_x + \sin\varphi \, \hat{e}_y ] \, , \\
 \rho_j &= \mbox{const}\, ,
\end{align}
\end{subequations}
where $\varphi$ parametrizes the orientation of the spin order in the $xy$-plane, and $M$ represents its amplitude.
N\'eel order has the period $P=2$, so that the matrix $\mathcal{H}^{(P)}_{\bk,\sigma}$ in Eq.~\eqref{eq: H_MF components} is two-dimensional and only the two parameters $\Delta_1^1$ and $\Delta_1^2$ contribute, where
\begin{subequations}
 \begin{align}
 & \Delta_1^1 = \Delta \cos \varphi \, , \\
 & \Delta_1^2 = \Delta \sin \varphi \, ,
 \end{align}
\end{subequations}
with $\Delta = UM$.

%%%%%%%%%%%%%%%%%%%%%%%%%%%%%%%%%%%%%%%%%%%%%%%%%%%%%%%%%%%%%%%%%%%%%%%%%%%%%%%

\subsubsection{Circular spiral order}

Circular spiral order has the form
\begin{subequations} \label{eq: spiral magn und charge}
\begin{align}
 \vec{S}_j &= M \big[ \cos(\bQ\!\cdot\!\mathbf{R}_j \!+\! \varphi) \, \hat{e}_x
 \pm \sin(\bQ\!\cdot\!\mathbf{R}_j \!+\! \varphi) \, \hat{e}_y \big] \, , \\
 \rho_j &= \mbox{const} \, ,
 \end{align}
\end{subequations}
where, as in the case of N\'eel order, $\varphi$ parametrizes the orientation of the spin order in the $xy$ plane and $M$ its (constant) amplitude. $\bQ$ is a generic wave vector of the form $(\pi-\delta,\pi)$ with $\delta > 0$. The "+" or "-" sign distinguishes between spirals rotating anti-clockwise and clockwise.

This type of order emerges as a special case of our general formalism if $\Delta_n^a$ is non-zero only for one mode $\bar{n}$ and its conjugate $P_S-\bar{n}$, with $\bar{n}$ odd and such that $\bQ$ can be approximated by $\bar{n}\bQ_P$ modulo a reciprocal lattice vector, with a suitably chosen $P$. Spiral order as in Eq.~\eqref{eq: spiral magn und charge} is then described by
\begin{subequations} \label{eq: spiral order}
\begin{align}
 \Delta_{\bar{n}}^1 &= \frac{1}{2} \Delta e^{i\varphi} \, , \quad
 \Delta_{P_S-\bar{n}}^1 = \frac{1}{2} \Delta e^{-i\varphi} \, ,  \\
 \Delta_{\bar{n}}^2 &= \mp \frac{i}{2} \Delta e^{i\varphi} \, , \quad
 \Delta_{P_S-\bar{n}}^2 = \pm \frac{i}{2} \Delta e^{-i\varphi} \, ,
\end{align}
\end{subequations}
with $\Delta = UM \in \mathbb{R}$. All other $\Delta_n^a$ are zero.
The matrix $\mathcal{H}^{(P)}_{\bk,\sigma}$ thus simplifies to a diagonal block matrix form with $P_S/2$ matrices of size two on the diagonal. Indeed spiral order can be described by a simpler $2\times 2$ mean-field Hamiltonian for each $\bk$-point, as previously used in mean-field calculations restriced to spiral states \cite{Chubukov1992, Chubukov1995, Dombre1990, Fresard1991, Igoshev2010, Yamase2016, Eberlein2016, Mitscherling2018, Bonetti2020a, Scholle2023}.
Note that the Eqs.~\eqref{eq: spiral order} apply only if $\bar{n} \neq P_S - \bar{n}$, which is fulfilled for any spiral state which is not a N\'eel state, that is, as long as $\bQ \neq (\pi,\pi)$.

%%%%%%%%%%%%%%%%%%%%%%%%%%%%%%%%%%%%%%%%%%%%%%%%%%%%%%%%%%%%%%%%%%%%%%%%%%%%%%%%%%%%%

\subsubsection{Stripe order}
We define as stripe order any type of collinear order that differs from N\'eel antiferromagnetism. In this case, the magnetization and charge densities have the form
\begin{subequations}
\begin{align}
 & \vec{S}_j = f_S(\mathbf{R}_j)
 [ \cos\varphi \,\hat{e}_x + \sin\varphi \,\hat{e}_y ] \, , \\
 & \rho_j = f_\rho(\mathbf{R}_j) \, ,
    \end{align}
\end{subequations}
where once again $\varphi$ is an angle parameterizing the orientation of the spins in the $xy$-plane, while $f_S(\mathbf{R}_j)$ and $f_\rho(\mathbf{R}_j)$ are two functions defining the spatial modulation of the magnetization amplitude and charge density, respectively. They can be expressed in terms of their Fourier coefficients as
\begin{subequations}
\begin{align}
 & f_S(\mathbf{R}_j) =
 \sum_{n\,\mathrm{odd}} M_n \, e^{in\bQ_P\cdot\mathbf{R}_j} \, , \\
 & f_\rho(\mathbf{R}_j) =
 \sum_{n\,\mathrm{even}} \rho_n \, e^{in\bQ_P\cdot\mathbf{R}_j}\, .
\end{align}
\end{subequations}

Stripe order can be obtained as a particular case of our general formalism, with $\Delta_n^a$ fulfilling
\begin{subequations}
\begin{align}
 & \Delta_n^0 = U \rho_n \, , \\
 & \Delta_n^1 = U M_n \cos \varphi \, , \\
 & \Delta_n^2 = U M_n \sin \varphi\, .
\end{align}
\end{subequations}
Depending on the coefficients $M_n$ and $\rho_n$, the spin and charge profiles can be sinusoidal or sharp (like domain walls) or anything in between.

%%%%%%%%%%%%%%%%%%%%%%%%%%%%%%%%%%%%%%%%%%%%%%%%%%%%%%%%%%%%%%%%%%%%%%%%%%%%%%

\subsection{Landau theory close to $T^*$}
\label{sec: Landau theory}

To determine the type of magnetic order close to the critical temperature $T^*$, we decouple the Hubbard interaction by introducing spin and charge order parameter fields via a Hubbard-Stratonovich transformation, and subsequently expand the resulting effective action in powers of the order parameters.
\subsubsection{Derivation of the effective action}
We write the Hubbard interaction as~\cite{Weng1991, Schulz1995, Borejsza2004}
\begin{equation} \label{eq: rewrite Hubb Int}
    \begin{split}
         U n_{j,\uparrow}n_{j,\downarrow} = \frac{U}{4} \left( c^\dagger_j c_j\right)^2 - \frac{U}{4} \left(c^\dagger_j \vec{\sigma}\cdot\hat{\Omega}_j c_j   \right)^2\, ,
    \end{split}
\end{equation}
where $\hat{\Omega}_j$ is an arbitrary site- and time-dependent unit vector. Intuitively, one can
imagine $\hat{\Omega}_j$ as being the direction of the local (both in space and time) magnetization. Because $\hat{\Omega}_j$ can be arbitrarily chosen, in the path integral of the Hubbard model we take the average over all possible $\hat{\Omega}_j$ with a properly defined measure such that $\int \mathcal{D}\hat{\Omega}=1$.
We cast the Hubbard interaction in the form~\eqref{eq: rewrite Hubb Int}, because this makes it compatible with our mean-field decoupling (see Eq.~\eqref{eq: MF decoupling}).

We perform a Hubbard-Stratonovich transformation to decouple each of the terms in Eq.~\eqref{eq: rewrite Hubb Int} by means of two fields, $\rho_j$ and $\rho^S_j$, representing fluctuations of the charge and spin amplitude, respectively.
Defining a spin field as $\vec{S}_j=\rho^S_j\,\hat{\Omega}_j$, we can represent the Hubbard interaction as
\begin{equation}
 e^{-U\int_0^\beta d\tau \sum_j n_{j\up} n_{j\down}} =
 \int \mathcal{D}\rho \, \mathcal{D}\vec{S}\,
 e^{-(\mathcal{S}_\rho + \mathcal{S}_S + \mathcal{S}_{\rm int})} \, ,
\end{equation}
where $\beta=1/T$ is the inverse temperature, and
\begin{subequations}
\begin{align}
 \mathcal{S}_\rho &= \frac{1}{U} \int_0^\beta\! d\tau\sum_j \rho_j^2\, , \\
 \mathcal{S}_{S} &= \frac{1}{U} \int_0^\beta\! d\tau\sum_j |\vec{S}_j|^2\, ,\\
 \mathcal{S}_{\rm int} &= \int_0^\beta \!d\tau \sum_j
 \bar{c}_j (i\rho_j +\vec{S}_j\cdot\vec{\sigma}) c_j \, .
\end{align}
\end{subequations}
To keep the notation light, we have dropped the time dependence of the bosonic
($\rho_j$, $\vec{S}_j$) and fermionic ($c_j$, $\bar{c}_j$) fields.

An effective action for the bosonic fields $\rho_j$ and $\vec{S}_j$ can be derived by integrating out the fermions. Except for a field-independent term, one obtains
\begin{equation} \label{eq: EffectiveBosonicAction}
\begin{split}
 &\mathcal{S}^{\text{eff}}\left[\rho,\vec{S}\right] =
 \frac{1}{U} \int_0^\beta d\tau\,\sum_j \left(\rho_j^2 + |\vec{S}_j|^2\right)  \\
 & \hskip1.6cm
 - \tr\ln\left[ \mathbb{1}_\mathrm{st}-G_0 \cdot \left(i\rho +\vec{\sigma}\cdot\vec{S}\right)
 \right] \, ,
\end{split}
\end{equation}
where $G_0$ is the Fourier transform to real space and imaginary time of the bare Matsubara Green's function $G_0(k) = (i\nu+\mu-\epsilon_\bk)^{-1}$, while $\rho$ and $\vec{S}$ are diagonal matrices in space and time defined as
$\rho_{jj'}(\tau,\tau') = \rho_j(\tau)\,\delta_{jj'}\delta(\tau\!-\!\tau')$ and
$\Vec{S}_{jj'}(\tau,\tau')=\vec{S}_j(\tau)\,\delta_{jj'}\delta(\tau\!-\!\tau')$.
The trace is summing over space and time indices, $G_0 A$ is the space-time matrix product
$\sum_{j^{\prime\prime}} \int_0^\beta d\tau^{\prime\prime} \,
 G_{0,jj^{\prime\prime}}(\tau\!-\!\tau^{\prime\prime}) \,
 A_{j^{\prime\prime}j'}(\tau^{\prime\prime},\tau')$, and
$\mathbb{1}_\mathrm{st} = \delta_{jj'} \delta(\tau-\tau')$ is the space-time unit matrix.

%%%%%%%%%%%%%%%%%%%%%%%%%%%%%%%%%%%%%%%%%%%%%%%%%%%%%%%%%%%%%%%%%%%%%%%%%%%%%%%%%%%%%

\subsubsection{Taylor expansion of the effective action}
We now expand the logarithm in Eq.~\eqref{eq: EffectiveBosonicAction} in powers of $\rho_j$ and $\vec{S}_j$. Such an expansion is justified in the vicinity of the critical temperature $T^*$, where the magnetic and charge order parameters are small.
To this end we write
\begin{equation}\label{eq: taylor series effective action}
\begin{split}
 \tr\ln\left[\mathbb{1}_\mathrm{st}-G_0 A\right] =
 \sum_{n=1}^{\infty}\frac{1}{n}\tr\left[\left(G_0A\right)^n\right]\, ,
\end{split}
\end{equation}
with $A = i\rho + \vec{\sigma}\cdot\Vec{S}$. The trace does not depend on the representation. In the following we perform all calculations in momentum and frequency space.

The quadratic term in $\vec{S}_q$, with $\vec{S}_q$ the (spatio-temporal) Fourier transform of $\vec{S}_j$, takes the form 
\begin{subequations}
\begin{align}
 & \int_q  \left[ U^{-1} - \Pi_0(q) \right] \vec{S}_{-q} \cdot \Vec{S}_q\, , \\
 & \Pi_0(q) =  -\int_k G_0(k) G_0(k+q)\, \label{eq: bare bubble},
\end{align}
\end{subequations}
$\int_k = T\sum_\nu\int_\bk$ is a shorthand for a sum over Matsubara frequencies and a momentum integration, and $q = (\bq,\Omega)$ is a collective variable comprising a lattice momentum and a bosonic Matsubara frequency. We define $T^*$ as the critical temperature where
$\mathrm{min}_\bq \big[ U^{-1} - \Pi_0(\bq,0) \big] = 0$, signaling an instability towards the formation of magnetic order. This condition is first met, in the most general case, at four symmetry related wave vectors in the Brillouin zone of the form $\bq = \pm \bQ_x$ or $\bq = \pm \bQ_y$, with $\bQ_x = (\pi-\delta, \pi)$ and $\bQ_y = (\pi, \pi-\delta)$. This means that the magnetic order forming right below $T^*$ can be entirely characterized by these wave vectors. Note that, because of the opposite sign between the two terms on the right hand side of Eq.~\eqref{eq: rewrite Hubb Int}, the coefficient of the term quadratic in $\rho_q$ is $U^{-1} +\Pi_0(q)$, which is always positive. Thus, within mean-field theory, an instability towards charge order alone can never occur in the Hubbard model.

For a mean-field study of the Taylor-expanded effective action~\eqref{eq: EffectiveBosonicAction}, we can therefore assume that $\vec{S}_q$ possesses solely modes at $\bq = \pm\bQ_x$ and $\bq = \pm\bQ_y$ close to $T = T^*$, corresponding to the ansatz
\begin{equation}\label{eq: S_Form}
\begin{split}
 \vec{S}_{(\bq,\Omega)} = &
  \Big[\vec{M}_x\, \delta(\bq-\bQ_x) + \vec{M}_x^* \,\delta(\bq+\bQ_x) \\
  &+ \vec{M}_y \,\delta(\bq-\bQ_y) + \vec{M}_{y}^*\, \delta(\bq+\bQ_y)\Big]
  \delta_{\Omega,0} \, ,
\end{split}
\end{equation}
where $\vec{M}_x$ and $\vec{M}_y$ are constant complex vectors. We assume static fields, consistent with our mean-field treatment.

Third order terms involving only spin fields vanish due to time-reversal symmetry.
The third order term involving two $\vec{S}_q$ fields and one $\rho_q$ field takes the form
\begin{equation}
 \int_{q,q'} i\lambda(q,q') \vec{S}_q \cdot \vec{S}_{-q'} \rho_{q'-q} \, ,
\end{equation}
with a coupling function $\lambda(q,q')$.
Inserting Eq.~\eqref{eq: S_Form} into this equation, we see that, within mean-field theory, the only charge modes that couple to $\vec{M}_x$ and $\vec{M}_y$ are those where $\bq = \bzero$, $\bq = \pm2\bQ_{x,y}$, and $\bq = \pm (\bQ_x \pm \bQ_y)$. Neglecting the $\bq = \bzero$ mode, which does not lead to any symmetry breaking, and higher order spin-charge interactions (this approximation will be justified below), we can write
\begin{equation} \label{rho_Form}
\begin{split}
 i\rho_{(\bq,\Omega)} =
 &\Big[\phi_x\, \delta(\bq-2\bQ_x) + \phi_x^{*} \, \delta(\bq+2\bQ_x) \\
 +&\phi_y\, \delta(\bq-2\bQ_y) + \phi_y^{*}\, \delta(\bq+2\bQ_y) \\
 +&\phi_+\, \delta(\bq-\bQ_+) + \phi_+^{*}\, \delta(\bq+\bQ_+) \\
 +&\phi_-\, \delta(\bq-\bQ_-) + \phi_-^{*}\, \delta(\bq+\bQ_-)\Big] \delta_{\Omega,0}\, ,
\end{split}
\end{equation}
where $\phi_x$, $\phi_y$, and $\phi_\pm$ are complex constants, and $\bQ_\pm = \bQ_x \pm \bQ_y$.

Inserting Eq.~\eqref{eq: S_Form} and \eqref{rho_Form} into Eq.~\eqref{eq: EffectiveBosonicAction}, and expanding up to quartic order in $\vec{M}_x$ and $\vec{M}_y$, to quadratic order in $\phi_x$, $\phi_y$, and $\phi_\pm$, and to third order in the mixed terms, we obtain the effective potential
\begin{equation} \label{eq: effective charge spin potential}
\begin{split}
 & V(\vec{M}_x,\vec{M}_y,\phi_x,\phi_y,\phi_\pm) =
 s \left( |\vec{M}_x|^2 +|\vec{M}_y|^2 \right) \\
 &+ u_0 \left(|\vec{M}_x|^2+|\vec{M}_y|^2\right)^2
  + u_1 \left(|\vec{M}_x|^2-|\vec{M}_y|^2\right)^2 \\
 &+ u_2 \left(|\vec{M}_x\cdot \vec{M}_x|^2 + |\vec{M}_y\cdot \vec{M}_y|^2\right) \\
 &+ u_3 \left(|\vec{M}_x\cdot \vec{M}_y|^2 + |\vec{M}_x\cdot \vec{M}_y^*|^2\right) \\
 &- r_1 \left(|\phi_x|^2+|\phi_y|^2\right) - r_2\left(|\phi_+|^2+|\phi_-|^2\right)\\
 &+ b_1 \left(\phi_x\,\vec{M}^*_x\cdot\vec{M}^*_x + \phi_y\,\vec{M}^*_y\cdot \vec{M}^*_y
 + \mathrm{c.c.} \right) \\
 &+ b_2 \left(\phi_+\,\vec{M}^*_x\cdot\vec{M}^*_y + \phi_-\,\vec{M}^*_x\cdot \vec{M}_y
 + \mathrm{c.c.} \right) \, .
\end{split}
\end{equation}
The coefficients $s$, $u_0$, $u_1$, $u_2$, $u_3$, $r_1$, $r_2$, $b_1$, and $b_2$ are determined by frequency and momentum integrals of products of bare propagators $G_0$. The concrete expressions are presented in Appendix~\ref{app: GLT coefficients}.

The charge degrees of freedom can be eliminated from the theory by imposing
$\partial V/\partial\phi_\alpha = 0$, for $\alpha = x,y,\pm$, which yields
\begin{subequations}\label{eq: charge bilinears}
\begin{align}
 & \phi_{x} = \frac{b_1}{r_1} \vec{M}_{x} \cdot \vec{M}_{x} \, , \quad
   \phi_{y} = \frac{b_1}{r_1} \vec{M}_{y} \cdot \vec{M}_{y} \, , \\
 & \phi_{+} = \frac{b_2}{r_2} \vec{M}_{x} \cdot \vec{M}_{y} \, , \quad
   \phi_{-} = \frac{b_2}{r_2} \vec{M}_{x} \cdot \vec{M}_{y}^* \, .
\end{align}
\end{subequations}
From the above equations we see that at the extremal points of the potential $V$ the charge order parameter is a bilinear of the spin order parameter. Therefore, in order to get an effective theory that is at most quartic in $\vec{M}_x$ and $\vec{M}_y$, one has to retain all and only the terms in Eq.~\eqref{eq: effective charge spin potential} in the expansion of the bosonic action~\eqref{eq: EffectiveBosonicAction}.
Inserting Eqs.~\eqref{eq: charge bilinears} into~\eqref{eq: effective charge spin potential}, one gets 
\begin{equation} \label{eq: effective spin potential}
\begin{split}
 & V_\mathrm{eff}(\vec{M}_x,\vec{M}_y) =
 s \left( |\vec{M}_x|^2 +|\vec{M}_y|^2 \right) \\
 &+ u_0 \left(|\vec{M}_x|^2+|\vec{M}_y|^2\right)^2
  + u_1 \left(|\vec{M}_x|^2-|\vec{M}_y|^2\right)^2 \\
 &+ \widetilde{u}_2 \left(
 |\vec{M}_x\cdot \vec{M}_x|^2 + |\vec{M}_y\cdot \vec{M}_y|^2 \right) \\
 &+ \widetilde{u}_3 \left(
 |\vec{M}_x\cdot \vec{M}_y|^2 + |\vec{M}_x\cdot \vec{M}_y^*|^2 \right) \ ,
\end{split}
\end{equation}
with $\widetilde{u}_2 = u_2 + b_1^2/r_1$ and $\widetilde{u}_3 = u_3 + b_2^2/r_2$.
Minimizing $V_\mathrm{eff}(\vec{M}_x,\vec{M}_y)$ with respect to $\vec{M}_x$ and $\vec{M}_y$ we can determine the magnetic state at temperatures right below $T^*$.
A Landau theory with an effective potential of the form~\eqref{eq: effective spin potential} has previously been derived from general symmetry arguments \cite{Zhang2002, DePrato2006, Sachdev2019}. The form of the Landau theory restricted to the case a single mode ($\vec{M}_x$ or $\vec{M}_y$) was derived earlier in Ref.~\cite{ZacharKivelson1998}.

%%%%%%%%%%%%%%%%%5%%%%%%%%%%%%%%%%%%%%%%%%%%%%%%%%%%%%%%%%%%%%%%%%%%%%%%%%%%%%%%%%%%%%

\subsection{Susceptiblities in the spiral state}
\label{sec: spiral susceptibilities}

With a proper redefinition of the local spin reference frame~\cite{Kampf1996,Bonetti2022}, spiral order as in Eq.~\eqref{eq: spiral magn und charge} can be described in terms of a $2\times2$ Hamiltonian of the form
\begin{equation}
 \mathcal{H}^\mathrm{sp}_\bk = \left(
 \begin{array}{cc}
 \epsilon_\bk & \Delta\,e^{-i\varphi} \\
 \Delta\,e^{i\varphi} & \epsilon_{\bk+\bQ}
 \end{array} \right) \, .
\end{equation}
Since the energy does not depend on the phase $\varphi$, we can choose, without loss of generality, $\varphi=0$. Within the rotated reference frame, one can compute the charge and spin susceptibilities within random phase approximation (RPA) as
\begin{equation}\label{eq: RPA spiral state}
 \chit(\bq,\omega) = \chit_0(\bq,\omega)
 \big[ \mathbb{1}_4 - \Gamma_0 \chit_0(\bq,\omega) \big]^{-1} \, ,
\end{equation}
where $\Gamma_0 = 2U \, \mathrm{diag}(-1,1,1,1)$, and the bare susceptibility $\chit_0(\bq,\omega)$, as a function of the bosonic Matsubara frequency $\Omega$, is given by
\begin{equation}\label{eq: spiral bubble}
 \chit^{ab}_0(\bq,i\Omega) = -\frac{1}{4} \int_k \mathrm{Tr} \left[
 \sigma^a G(\bk+\bq,\nu+\Omega) \, \sigma^b G(\bk,\nu) \right] ,
\end{equation}
where $G(\bk,\nu) = \big[ (i\nu+\mu)\mathbb{1}_2-\mathcal{H}^\mathrm{sp}_\bk \big]^{-1}$ is the mean-field Green's function. The real frequency susceptibility is obtained by substituting $i\Omega \to \omega+i0^+$ after performing the Matsubara sum.

To compute the susceptibilities in the physical (unrotated) spin reference frame, one must rotate Eq.~\eqref{eq: RPA spiral state} in the $xy$ plane with a spatially dependent angle of $\bQ\cdot\mathbf{R}_j$~\cite{Kampf1996,Bonetti2022}. Such a rotation will produce in general momentum off-diagonal components of the susceptibilities, as spiral order breaks translational invariance. However, for our purpose of a stability analysis of the spiral state it suffices to consider the susceptibilities in the rotated spin reference frame.

%%%%%%%%%%%%%%%%%%%%%%%%%%%%%%%%%%%%%%%%%%%%%%%%%%%%%%%%%%%%%%%%%%%%%%%%%%%%%%%%

\section{Results} \label{sec: results}

We now present the mean-field phase diagram of the two-dimensional Hubbard model as obtained from the three complementary methods described in the preceding section.
We choose a sizable next-nearest neighbor hopping $t' = -0.3t$, as is frequently used to model the band structure and Fermi surface of the cuprate superconductor yttrium barium copper oxide (YBCO) \cite{Pavarini2001}. In the hole-doped region ($n < 1$) we obtain the same sequence of magnetic states also for other negative values of $t'/t$.
For the interaction strength we choose $U = 3t$, which is strong enough to obtain magnetic order in spite of the magnetic frustration imposed by $t'$, but weak enough to obtain qualitatively plausible results from the Hartree-Fock approximation. In the cuprates the Hubbard interaction is much larger, but the effective interaction driving magnetic order and magnetic correlations is renormalized to smaller values by fluctuations.

Since the magnetic order at densities $n \geq 1$ is generally of N\'eel type \cite{Scholle2023}, we focus on the {\em hole-doped}\/ regime $n < 1$, where an intriguing sequence of ordering patterns is found.

%%%%%%%%%%%%%%%%%%%%%%%%%%%%%%%%%%%%%%%%%%%%%%%%%%%%%%%%%%%%%%%%%%%%%%%%%%%%%%%%%%%

\subsection{Ground state phase diagram} \label{sec: groundstate}

\begin{figure}[t!]
\centering
\includegraphics[width=0.5\textwidth]{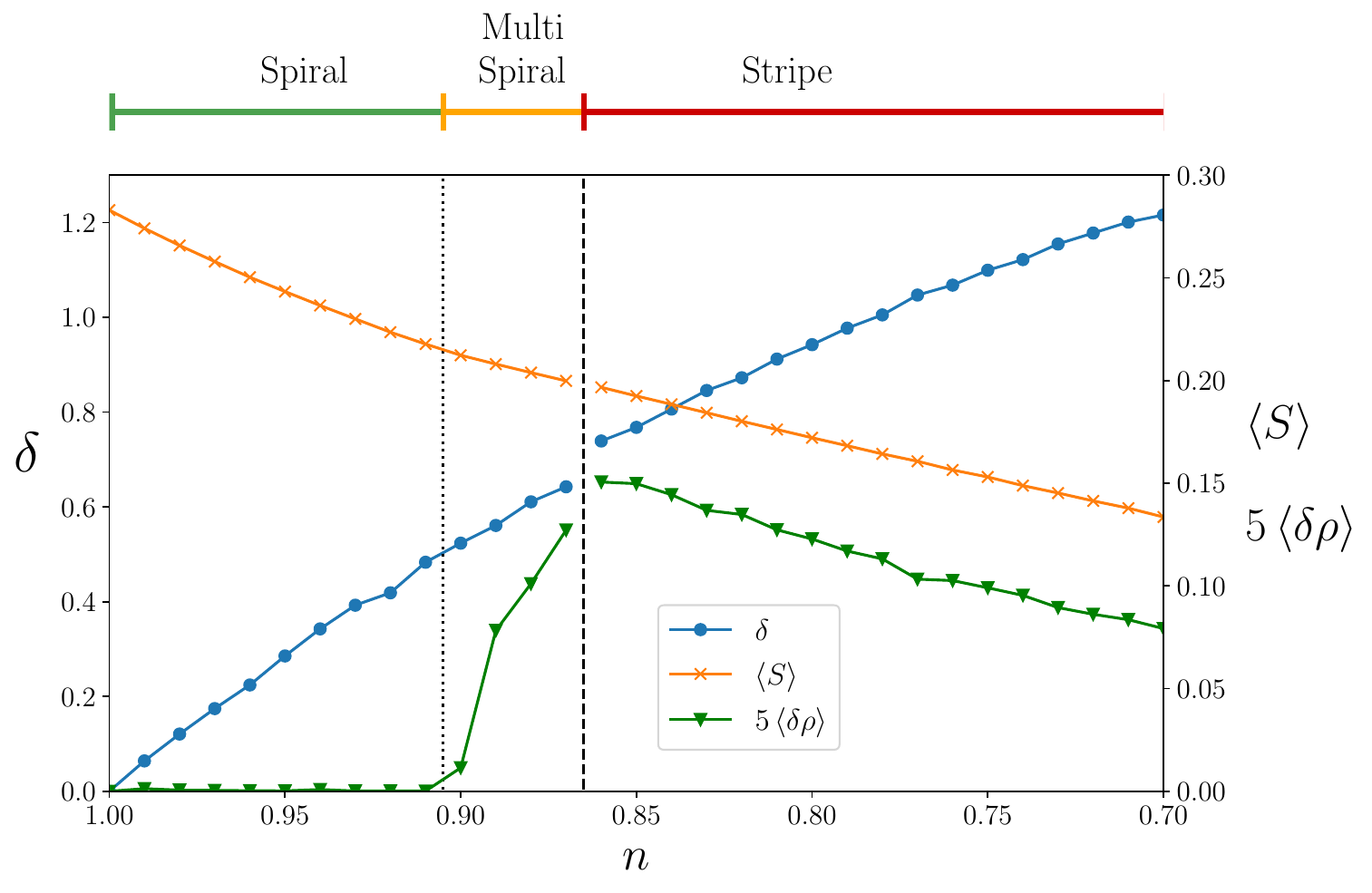}
 \caption{Top: Magnetic phases in the ground state of the 2D Hubbard model with $t'=-0.3t$ and $U=3t$, for densities $0.7 \leq n < 1$. Bottom: Incommensurability $\delta$, average spin amplitude $\langle S \rangle$, and average charge modulation $\langle\delta\rho\rangle$ as functions of $n$.}
\label{fig: GS phase diag}
\end{figure}
In Fig.~\ref{fig: GS phase diag} we show the ground state phase diagram as obtained from the mean-field solution described in Sec.~\ref{sec: MFT}.
The N\'eel state at half-filling ($n=1$) is immediately unstable toward a spiral state upon hole doping.
At $n \simeq 0.9$, the spiral state becomes unstable, leading into a more complex phase at lower densities which is still coplanar and non-collinear, but with a modulated spin amplitude and charge density. We discuss this phase, which we call {\em multi-spiral}, in more detail in Sec.~\ref{sec: multispiral}.
Upon further increasing the hole-doping, a conventional stripe state with collinear spin order and charge density wave order is stabilized.
The transitions from the N\'eel to the spiral state and from the spiral to the multi-spiral state are continuous, while the transition from the multi-spiral state to the stripe state might be first order.

For a quantitative characterization of the various states, we define the average spin amplitude $\langle S \rangle$ and the average charge modulation $\langle \delta\rho\rangle$ as
\begin{subequations}
\begin{align}
 \langle S \rangle &=
 \sqrt{ P_S^{-1} \sum_{j\in\mbox{cell}} |\vec{S}_j|^2 } \, , \\
 \langle \delta\rho\rangle &=
 \sqrt{ P_S^{-1} \sum_{j\in\mbox{cell}} |\rho_j-n|^2 } \, ,
\end{align}
\end{subequations}
where the lattice sum extends over one unit cell (with $P_S$ sites). The dominant wave vector is parametrized by the incommensurability
\begin{equation}
 \delta = \pi - Q_x^{\mathrm{max}} \, ,
\end{equation}
with $\bQ^\mathrm{max} = n_\mathrm{max} \bQ_P$, where $n_\mathrm{max}$ is the index belonging to the largest magnetic gaps $\Delta_n^1$ or $\Delta_n^2$. In Fig.~\ref{fig: GS phase diag} we see that, irrespective of the phase transitions occurring, both $\langle S \rangle$ and $\delta$ display a rather smooth and monotonic behavior.
By contrast, $\langle\delta\rho\rangle$ vanishes in the N\'eel and spiral phases, but then rises quickly in the multi-spiral regime, peaking at the transition to the stripe phase. It then slowly decays as the density is further decreased.

%%%%%%%%%%%%%%%%%%%%%%%%%%%%%%%%%%%%%%%%%%%%%%%%%%%%%%%%%%%%%%%%%%%%%%%%%%%%%%%%

\subsection{Phase diagram at $T=T^*$} \label{sec: T*phasediagram}

\begin{figure}[t!]
\centering
 \includegraphics[width=0.5\textwidth]{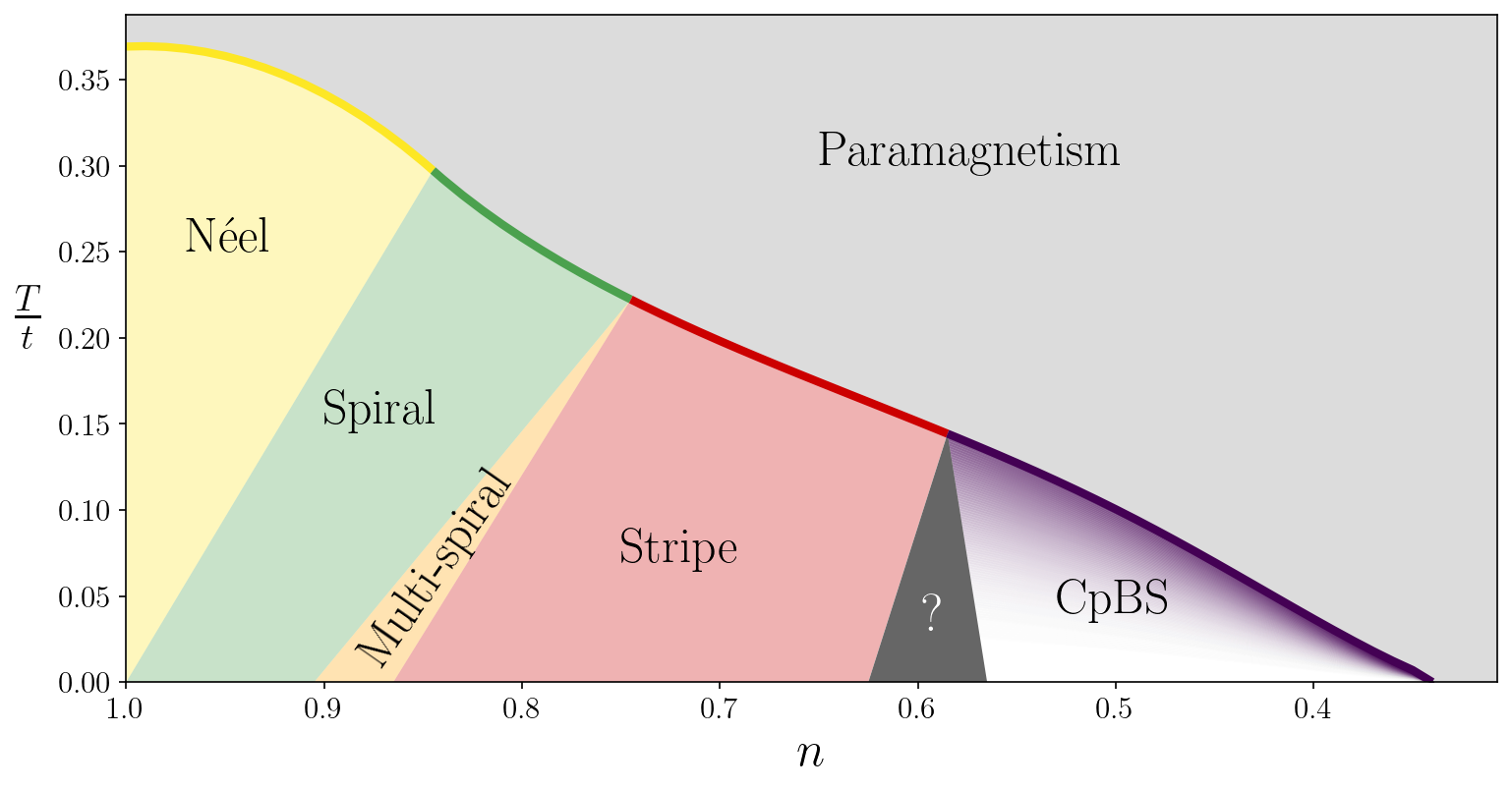}
 \caption{Mean field phase diagram of the Hubbard model for $t^\prime = -0.3t$ and $U=3t$. The various types of magnetic order appearing at $T = T^*$ were determined from Landau theory, the states at $T = 0$ by solving the Hartree-Fock equations derived in Sec.~\ref{sec: MFT}. The phase transitions for $0 < T < T^*$ are indicated only schematically by straight lines connecting the calculated transition points at $T=T^*$ and $T=0$.}
\label{fig: T* phase diag}
\end{figure}
In Fig.~\ref{fig: T* phase diag} we show the various magnetic phases we obtain at $T=T^*$ by minimizing the effective potential~\eqref{eq: effective spin potential}. In a density regime near half-filling, at $T=T^*$ the maximum of $\Pi_0(\bq,0)$ in Eq.~\eqref{eq: bare bubble} occurs at $\bq=(\pi,\pi)$, implying that the phase being realized right below the (mean-field) critical temperature is a N\'eel antiferromagnet. Reducing the density, $\Pi_0(\bq,0)$ develops four identical maxima at $\bq = \pm\bQ_x$ and $\pm\bQ_y$ at $T^*$. An analysis of the quartic terms in the effective potential \eqref{eq: effective spin potential} reveals that the N\'eel state is replaced by a spiral phase characterized by
\begin{subequations}\label{eq: spiral in Landau th}
\begin{align}
 & \vec{M}_x = \frac{M}{\sqrt{2}} \left(\hat{e}_1 \pm i\hat{e}_2\right) \, , \\
 & \vec{M}_y = 0 \, ,
\end{align}
\end{subequations}
or by the same expression with $\vec{M}_x \leftrightarrow \vec{M}_y$.
Here, $\hat{e}_1$ and $\hat{e}_2$ are two orthogonal real unit vectors. The spiral phase maintains a uniform charge density, but it breaks the $C_4$ rotational symmetry of the square lattice.

At larger hole dopings, the spiral phase is replaced by a stripe phase, such that
\begin{subequations}\label{eq: stripe in Landau th}
\begin{align}
 \vec{M}_x &= M e^{i\varphi}\, \hat{e} \, , \\
 \vec{M}_y &= 0 \, ,
\end{align}
\end{subequations}
with an arbitrary unit vector $\hat{e}$, or by the same expression with $\vec{M}_x$ and $\vec{M}_y$ interchanged. This phase displays collinear magnetic order, a modulation of the charge density $\rho_j-n\propto \cos(2\bQ_x+2\varphi)$, and it breaks the $C_4$ symmetry.

A smooth interpolation between (circular) spiral and stripe order is given by {\em elliptical spiral} order \cite{ZacharKivelson1998},
\begin{subequations}\label{eq: elliptical spiral}
\begin{align}
 & \Vec{S}_j = M \left(
 \cos\alpha \cos\phi_j \, \hat{e}_1 \pm \sin\alpha \sin\phi_j \, \hat{e}_2 \right) \, , \\
 & \rho_j - n \propto \cos(2\alpha) \cos(2\phi_j)\, ,
\end{align}
\end{subequations}
with $\phi_j = \bQ_x\cdot\mathbf{R}_j+\varphi$ or $\bQ_y\cdot\mathbf{R}_j+\varphi$. The parameter $\alpha$ allows for a smooth interpolation between a spiral ($\alpha=\pi/4$) and a stripe ($\alpha=0$) phase. At the transition point between spiral and stripe order, the effective potential \eqref{eq: effective spin potential} is degenerate with respect to variations of $\alpha$. For $T < T^*$, this degeneracy is lifted by higher order terms (beyond quartic).

At even lower densities, we find a coplanar bidirectional stripe phase (CpBS), characterized by
\begin{subequations}
\begin{align}
 \vec{M}_x &= M e^{i\varphi_1}\, \hat{e}_1 \, , \\
 \vec{M}_y &= M e^{i\varphi_2}\, \hat{e}_2\, ,
\end{align}
\end{subequations}
with orthogonal unit vectors $\hat{e}_1$ and $\hat{e}_2$, and arbitrary phases $\varphi_1$ and $\varphi_2$. The charge density is then modulated as $\rho_j-n \propto \cos(2\bQ_x\cdot\mathbf{R}_j+2\varphi_1)+\cos(2\bQ_y\cdot\mathbf{R}_j+2\varphi_2)$. It is therefore conceivable that, at a finite distance below the $T=T^*$ line, a new phase emerges between unidirectional stripe and CpBS orders, interpolating between the two. We have marked this possible intermediate phase with a question mark in Fig.~\ref{fig: T* phase diag}.
The white color in the low temperature regime of the CpBS phase indicates that we have not clarified the nature of this phase far below $T^*$. More complex ordering patterns are possible there \cite{Scholle2023}, but in this regime of very large hole doping any magnetic order is probably an artifact of mean-field theory, and thus of limited interest.

%%%%%%%%%%%%%%%%%%%%%%%%%%%%%%%%%%%%%%%%%%%%%%%%%%%%%%%%%%%%%%%%%%%%%%%%%%%%

\subsection{Instability of the spiral state} \label{sec: multispiral}

The spiral state is stable in a finite hole-doping range near half-filling. At larger hole-doping, collinear stripe states have the lowest energy. We now clarify the nature of the instability of the spiral state upon increasing hole-doping, and the transition to a stripe state.
The instability of the spiral state can be detected by analyzing the static charge and spin susceptibilities. At zero frequency, the bare susceptibilities $\tilde\chi_0^{ab}$ in Eq.~\eqref{eq: spiral bubble} vanish if $a=3$ and $b\neq 3$ or $a\neq3$ and $b=3$.
Hence, the $a,b=0,1,2$ sector of the susceptibilities, corresponding to charge, spin amplitude, and in-plane spin orientation fluctuations, decouples from the $a,b=3$ sector, which is associated with out-of-plane spin orientation fluctuations.
An instability is signaled by a divergence and subsequent sign change of the susceptibilities in Eq.~\eqref{eq: RPA spiral state}. Such a divergence must however be distinguished from divergences due to Goldstone modes. Within our conventions, the Goldstone modes of the spiral state manifest themselves as $\chit^{22}(\bzero,0) = \infty$ and $\chit^{33}(\pm\bQ,0) = \infty$ in the rotated spin frame \cite{Kampf1996, Bonetti2022}.
The static out-of-plane spin susceptibility $\chit^{33}(\bq,0)$ remains always positive and finite for $\bq \neq \pm\bQ$.

We therefore search for a diverging susceptibility in the $a,b=0,1,2$ sector at $\bq \neq \bzero$, which is necessarily associated with an eigenvalue of the RPA denominator in Eq.~\eqref{eq: RPA spiral state} crossing zero. Hence, to determine the instability of the spiral state, and the nature of the magnetic order beyond the instability line, we study the eigenvalues of the matrix
\begin{equation}
 D(\bq) = \left(
 \begin{array}{ccc}
 - \frac{1}{2U} - \chit_0^{00}(\bq) & - \chit_0^{01}(\bq) & - \chit_0^{02}(\bq) \\
 - \chit_0^{10}(\bq) & \frac{1}{2U} - \chit_0^{11}(\bq) & - \chit_0^{12}(\bq) \\
 - \chit_0^{20}(\bq) & - \chit_0^{21}(\bq) & \frac{1}{2U} - \chit_0^{22}(\bq)
\end{array}
\right) \, ,
\end{equation}
with $\chit_0^{ab}(\bq) = \chit_0^{ab}(\bq,0)$. The spiral state is stable if the matrix has two positive and one negative eigenvalues for all $\bq \neq \bzero$, and viceversa when it is unstable.

\begin{figure}[t!]
\centering
\includegraphics[width=0.475\textwidth]{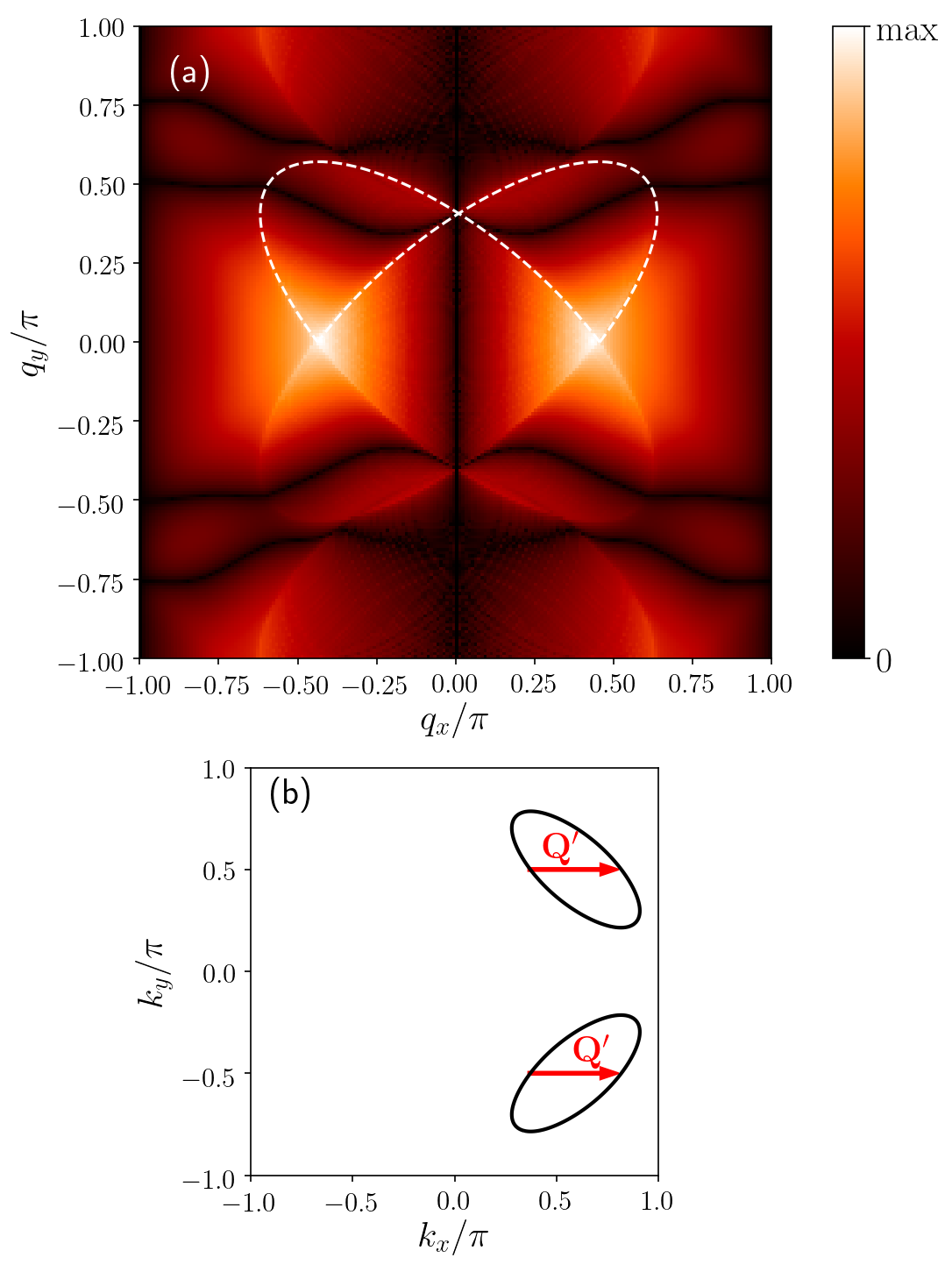}
 \caption{Panel~(a): Momentum dependence of $|\chit_0^{12}(\bq)|$, displaying nonanalyticities  on the two so-called $2k_F$-lines. The absolute maxima occur where these two lines cross on the $q_x$ axis. In the upper half of the Brillouin zone ($q_y>0$), the $2k_F$-lines are retraced by dashed white lines as a guide to the eye.
 Panel~(b):~Fermi surface in the spiral state, consisting of two hole pockets. The scattering processes with a momentum transfer of $\bQ'$, connecting opposite sides of the pockets with parallel tangents, are highlighted by red arrows. \\
 Parameters: $T=0$, $t'=-0.3t$, $n=0.90$, $U=3t$, for which mean-field theory yields $\Delta \approx 0.637t$ and $\delta \approx 0.167\pi$.}
\label{fig: 2kF line}
\end{figure}

We define $\bQ'$ as the non-zero wave vector at which the absolute value of the second largest eigenvalue of $D(\bq)$ has a global minimum. With some lengthy but straightforward algebra, one can prove that $\bQ'$ is the momentum at which $|\chit_0^{12}(\bq)|$ is maximal. In the ground state, $\bQ'$ is entirely determined by the geometry of the Fermi surface in the spiral state, which, at least for small dopings, consists of two hole pockets centered at $(\frac{\pi+\delta}{2},\pm\frac{\pi}{2})$ with $\delta > 0$, see panel (b) of Fig.~\ref{fig: 2kF line}.
\begin{figure*}[t!]
\centering
\includegraphics[width=\textwidth]{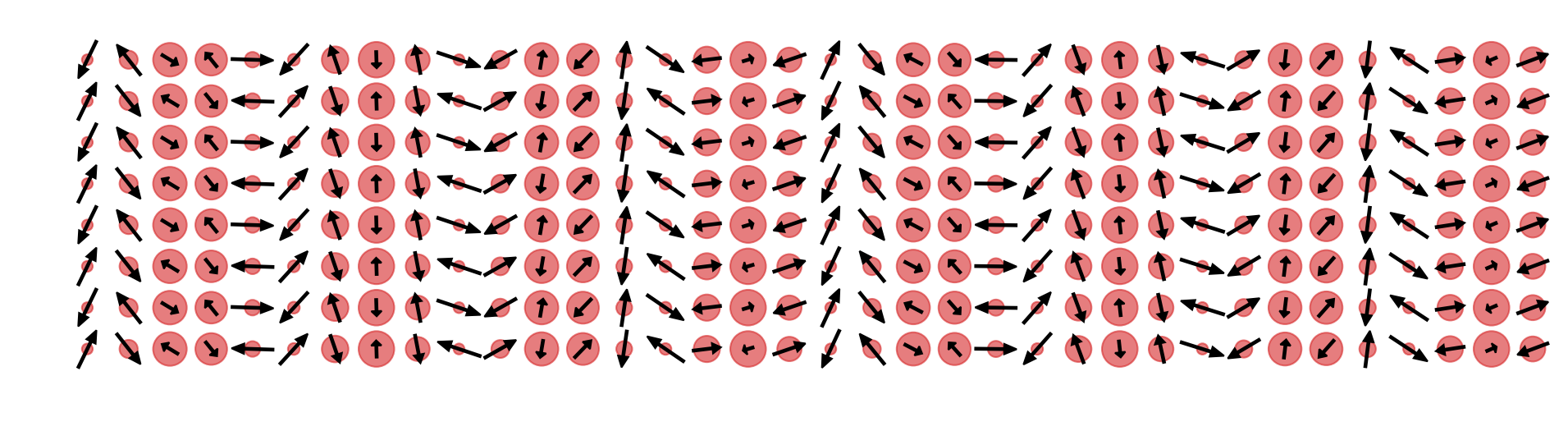}
 \caption{Spin and charge order pattern for a 3Q spiral state. Bigger (smaller) red bubbles represent a higher (lower) local hole concentration. The black arrows represent the local magnetization vector $\vec{S}_j$. The parameters are the same as in Tab.~\ref{tab: 3Q spiral MF vs RPA}, except for $M'$, which we have enhanced for visualization purposes.}
\label{fig: 3Q spiral pattern}
\end{figure*}
In panel (a) of Fig.~\ref{fig: 2kF line} we see that $|\chit_0^{12}(\bq)|$ has pronounced peaks at crossing points of two ellipse-shaped lines in $\bq$ space, on which $|\chit_0^{12}(\bq)|$ exhibits a singularity. These lines are ``$2k_F$-lines'' \cite{Holder2012}  corresponding to the set of wave vectors connecting points with parallel tangents on the Fermi surfaces of the hole pockets. They can be geometrically constructed by shifting the two hole pockets such that their centers coincide with the $\Gamma$ point $(0,0)$, and rescaling them by a factor of two. The global maximum of $|\chit_0^{12}(\bq)|$ occurs where the two $2k_F$-lines cross. As displayed in Fig.~\ref{fig: 2kF line}, there are two pairs of crossings, one occurring on the $q_x$ axis ($q_y=0$), and one on the $q_y$ axis ($q_x=0$). Since $\chit_0^{12}(\bq)$ is identically zero along the $q_y$ axis \cite{Bonetti2022}, $\bQ'$ and $-\bQ'$ are determined as the points in momentum space where the two $2k_F$-lines cross on the $q_x$ axis. Using these prescriptions, an analytical expression for $\bQ' = (q',0)$ can be derived:
\begin{equation}
 q' = 2 \mathrm{arccos} \left[ \tilde{\mu}\sin(\delta/2) +
 \sqrt{\tilde{\Delta}^2 + \left( 1-\tilde{\mu}^2 \right) \cos^2(\delta/2)} \, \right] ,
\end{equation}
where $\Tilde{\mu} = \mu/(2t)$ and $\Tilde{\Delta} = \Delta/(2t)$.

The eigenvector of $D(\bq)$ corresponding to the smallest positive eigenvalue (in the regime of stability of spirals) or the largest negative one (in the regime of instability of spirals) can be shown to take the general form $(\varrho_0,1,i\gamma)$, with $\varrho_0,\gamma \in \mathbb{R}$. This form can be deduced by using that $\chit_0^{12}(\bq) = -\chit_0^{21}(\bq)$ and $\chit_0^{02}(\bq) = -\chit_0^{20}(\bq)$ are purely imaginary, while all other entries of $D(\bq)$ are purely real~\cite{Bonetti2022}. The form of the eigenvector corresponding to the eigenvalue of $D(\bq)$ that can cross zero enables us to derive the form of the magnetic and charge ordering occurring right beyond the instability line. In the rotated frame in which spiral order appears as ferromagnetic, the order parameters take the form
\begin{subequations}\label{eq: ell spiral OPs in rotated frame}
\begin{align}
 & \vec{\widetilde{S}}_j =
 M \left( \begin{array}{c} 1 \\ 0 \\ 0 \end{array} \right) +
 M' \left( \begin{array}{c}
 \cos(\bQ' \cdot \mathbf{R}_j + \varphi') \\
 \gamma \sin(\bQ' \cdot \mathbf{R}_j + \varphi') \\
 0 \end{array} \right) \, , \\
 & \delta\rho_j = M' \varrho_0 \cos(\bQ' \cdot \mathbf{R}_j + \varphi')\, ,
\end{align}
\end{subequations}
where $M'$ is an overall amplitude and $\varphi'$ a phase. Assuming an anti-clockwise rotating spiral proportional to $\cos(\bQ\cdot\mathbf{R}_j)\hat{e}_x+ \sin(\bQ\cdot\mathbf{R}_j)\hat{e}_y$, corresponding to $\gamma>0$, rotating Eqs.~\eqref{eq: ell spiral OPs in rotated frame} to the physical spin reference frame yields
\begin{equation} \label{eq: 3Q spiral}
 \vec{S}_j = \left(
 \begin{array}{c}
 M \cos\phi_j^\mathrm{sp} + M'_{+} \cos\phi_j^{+} + M'_{-} \cos\phi_j^{-} \\
 M \sin\phi_j^\mathrm{sp} - M'_{+} \sin\phi_j^{+} + M'_{-} \sin\phi_j^{-} \\
 0
 \end{array} \right) ,
\end{equation}
where $\phi_j^\mathrm{sp} = \bQ\cdot\mathbf{R}_j$,
$\phi_j^\pm = \bQ'_\pm \cdot \mathbf{R}_j -\varphi'$ with $\bQ'_\pm = -(\bQ' \pm \bQ)$, and $M'_{\pm} = M' (1 \pm \gamma)/2$. The charge order parameter is left unchanged by the rotation. Eq.~\eqref{eq: 3Q spiral} describes a magnetic state with three overlapping spirals with distinct wave vectors, two of which propagate anti-clockwise (those with $\bQ$ and $\bQ'_{-}$), and one clockwise (with $\bQ'_{+}$). Thus, one can label this state as 3Q spiral. Such a state is found also by our numerical calculations using the formalism discussed in Sec.~\ref{sec: MFT}.
\begin{table}[b]
\centering
\begin{tabular}{c||c|c|c|c|c|c|c}
 & $Q$ & $Q'_{+}$ & $Q'_{-}$ & $Q'$ & $\gamma$ & $\varrho_0$ & $M'$ \\
 \hline \hline
 $D(\bq)$ & 2.616 & 2.269 & 1.218 & 1.398 & 1.568 & 0.258 & - \\ \hline
 MF & 2.618& 2.269 & 1.222 & 1.396 & 1.567 & 0.261 &0.006
\end{tabular}
 \caption{Comparison of the values of $\bQ = (Q,\pi)$, $\bQ'_- = (Q'_-,\pi)$, $\bQ'_+ = (Q'_+,\pi)$, $\bQ' = (Q',0)$, $\gamma$, $\varrho_0$, and $M'$ (see text) as predicted from calculations in the spiral phase (first row) and as numerically obtained from the mean-field theory of Sec.~\ref{sec: MFT} (second row). }
\label{tab: 3Q spiral MF vs RPA}
\end{table}
In Table~\ref{tab: 3Q spiral MF vs RPA}, we report the ground state values of $\bQ$, $\bQ'_\pm$, $\bQ'$, $\gamma$, and $\rho_0$ as predicted by the analysis of the susceptibilities in the spiral state, and as computed by solving the mean-field equations from \ref{sec: MFT} for $n=0.90$, that is, right beyond the instability line (see Fig.~\ref{fig: GS phase diag} for comparison). The lowest energy state was found to have a period of $P=36$.
In Fig.~\ref{fig: 3Q spiral pattern}, we show the spin and charge patterns for a 3Q spiral state.

If the strength of the spiral order is weakened by raising the temperature, $\chit^{12}_0(\bq)$ approaches $[\Pi_0(\bq+\bQ,0) - \Pi_0(\bq-\bQ,0)]/(2i)$, with $\Pi_0(q)$ the bare bubble defined as in Eq.~\eqref{eq: bare bubble}. At the onset of magnetic order $\Pi_0(\bq,0)$ is peaked exactly at $\bQ$ (and symmetry related), which implies that for very weak $\Delta$ one has $\bQ' = -2\bQ$ modulo a reciprocal lattice vector. Similarly, in this limit one observes that $\chit_0^{01}(\bq)$, and $\chit_0^{02}(\bq)$ become zero, and that $\chit_0^{11}(\bq)$ and $\chit_0^{22}(\bq)$ approach the same value. For this reason the eigenvector of $D(\bq)$ that can cross zero takes the form $(0,1,i)$ in the limit of vanishing spiral order, that is, $\gamma \to 1$. This, together with $\bQ' \to -2\bQ$, implies that Eq.~\eqref{eq: 3Q spiral} takes the form of an elliptical spiral (see Eq.~\eqref{eq: elliptical spiral}).
\begin{figure}[t!]
\centering
\includegraphics[width=0.4\textwidth]{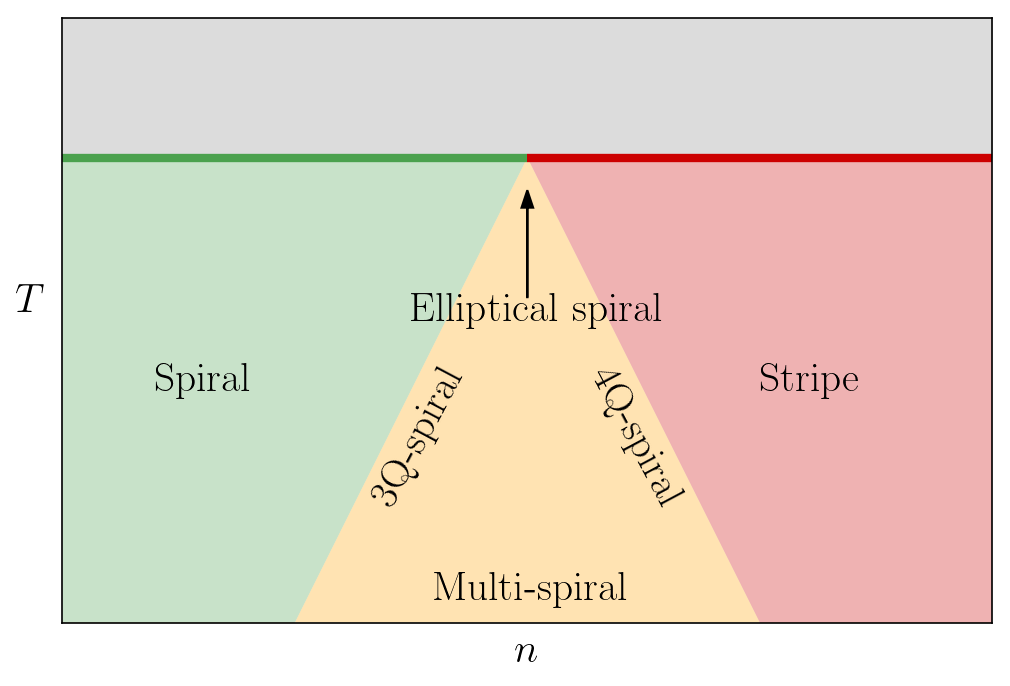}
 \caption{Schematic phase diagram of the mean-field transition from spiral to stripe order. We call in general multi-spiral the intermediate phase between spiral and stripe order. Closer to the spiral phase, we find three contributing Q-vectors (Eq.~\eqref{eq: 3Q spiral}), while close to the stripe phase, we observe four (Eq.~\eqref{eq: 4Q spiral}). Close to the critical temperature, the multi-spiral phase asymptotically approaches elliptical spiral order with only one Q-vector.}
\label{fig: schematic PD}
\end{figure}
Thus, the multi-spiral phase smoothly turns into an elliptical spiral phase as one raises the temperature toward $T^*$, as schematically shown in Fig.~\ref{fig: schematic PD}.

When moving away from the instability line of the spiral phase by increasing the doping, we find a fourth mode emerging, such that the spin order assumes the form
\begin{equation}\label{eq: 4Q spiral}
 \Vec{S}_j = \sum_{n=1}^4 M_n \left[ \cos(\bQ_n\!\cdot\!\mathbf{R}_j) \hat{e}_x +
 (-1)^n \sin(\bQ_n\!\cdot\!\mathbf{R}_j) \hat{e}_y \right] \, ,
\end{equation}
where we have dropped possible phases in the sine and cosine functions, and $\bQ_1 = \bQ$, $\bQ_2 = \bQ'_{+}$, $\bQ_3 = \bQ'_{-}$. We also observe, upon increasing doping, that $M_2 \to M_1$, $\bQ_2 \to \bQ_1$, $M_4 \to M_3$, and $\bQ_4 \to \bQ_3$. Hence, $S_j$ in Eq.~\eqref{eq: 4Q spiral} could gradually turn into a collinear stripe order with two harmonics:
\begin{equation}
 \Vec{S}_j = 2 \big[ M_1 \cos(\bQ_1\cdot\mathbf{R}_j) +
 M_3 \cos(\bQ_3\cdot\mathbf{R}_j) \big] \hat{e}_x \, .
\end{equation}
At the lowest density $n=0.87$ evaluated numerically in the multi-spiral regime, we find $M_2 \approx 0.5 M_1$, while at $n=0.86$ we already find a stripe phase with a single Q-vector. Hence, either $M_2$ grows to approach $M_1$ very quickly in a small density range, or the transition from multi-spiral to stripe is of first order. This is the reason why we have interrupted the lines that serve as a guide to the eye in Fig.~\ref{fig: GS phase diag} at the transition point.

%%%%%%%%%%%%%%%%%%%%%%%%%%%%%%%%%%%%%%%%%%%%%%%%%%%%%%%%%%%%%%%%%%%%%%%%%%%%%

\section{Conclusion} \label{sec: Conclusion}

Complementing our previous real-space Hartree-Fock study \cite{Scholle2023} by various additional techniques, we have obtained an almost complete understanding of the mean-field phase diagram of the two-dimensional Hubbard model with a moderate interaction strength. A large variety of distinct magnetic states appears, some with and some without concomitant charge order.
Since, in presence of a sizable next-nearest neighbor hopping, the magnetic states in the electron doped regime (filling $n > 1$) are always N\'eel ordered \cite{Scholle2023}, we focused on the hole doped regime $n < 1$.

The analysis in Ref.~\cite{Scholle2023} showed that the magnetic order of the Hubbard model is always coplanar and usually unidirectional, with wave vectors of the form $(\pi-\delta,\pi)$. Bidirectional order was found only at very small densities (large hole doping). Allowing for arbitrary coplanar and unidirectional order, we were able to solve the mean-field equations directly in the thermodynamic limit.
In the ground state, we thereby confirmed the circular spiral order at low hole-doping and the stripe order at large hole-doping. In between, we discovered a new {\em multi-spiral}\/ phase consisting of a superposition of various spirals with distinct but unidirectional wave vectors. Unlike the single-component spiral phase, the multi-spiral phase exhibits charge order similar to the stripe phase. Analyzing the spin-charge susceptibility of the spiral phase at its instability point, we found that the additional wave vectors contributing to the multi-spiral phase are related to nesting vectors of the hole-pockets in the simple spiral state.
We complemented the ground state calculation by a Landau free-energy analysis of the magnetic states right below the mean-field transition temperature $T^*$, where we found the following sequence of states as a function of increasing hole doping: N\'eel -- circular spiral -- unidirectional stripe -- bidirectional stripe. The multi-spiral phase found in the ground state becomes narrower (in doping) upon increasing temperature, and collapses to a point at $T^*$. Approaching that point from below ($T < T^*$), the multi-spiral state converges to an elliptical spiral with a single wave vector.

Our results are thus largely consistent with the previous real space Hartree-Fock calculation on large but finite lattices \cite{Scholle2023}. Only the multi-spiral phase could not be identified in the real space calculation, since the superposition of three or more contributing wave vectors leads naturally to very large unit cells.

To keep our paper concise, we fixed the next-nearest neighbor hopping and the Hubbard interaction to one value, $t'=-0.3t$ and $U=3t$, respectively, in all numerical results. The nearest-neighbor hopping $t$ sets the global energy scale. Qualitative changes of the phase diagram upon changing these parameters can be described as follows. Setting $t'=0$, the phase diagram becomes electron-hole symmetric. The magnetic order of the ground states is either N\'eel (at half-filling), or stripe (away from half-filling). Spiral states near half-filling exist only at finite temperatures in this special case \cite{Scholle2023}. From a continuity argument it is clear that for very small finite $t'$, spiral and stripe order will still be present also in the electron doped regime. However, already for $t'/t = -0.15$, the electron-doped regime is exclusively N\'eel ordered for $U=3t$ \cite{Scholle2023}.
Bidirectional stripe order at large hole doping appears only for a rather large $t'$. It is absent for $t'/t = -0.15$ \cite{Scholle2023}.
Decreasing $U$ obviously reduces the magnetically ordered regime, both in density and temperature. For sufficiently weak $U$ and a negative $t'/t$, a N\'eel ground state can be stable even for (small) finite hole doping \cite{Chubukov1992,Yamase2016}.
For $t'=0$ there is magnetic order at and near half-filling for any non-zero $U$, due to perfect nesting of the half-filled Fermi surface, while for $t' \neq 0$ a certain minimal interaction strength is required.

We finally discuss how order parameter fluctuations affect the phase diagram.
The Mermin-Wagner theorem \cite{Mermin1966} dictates that the spin SU(2) symmetry cannot be broken at any finite temperature. Hence, magnetic long-range order appearing in mean-field theory is destroyed by order parameter fluctuations.
However, some secondary order parameters emerging in the phases described above may survive in the form of vestigial order. Moreover, features of the spectral function for fermionic single-particle excitations in the magnetically ordered regime, such as the Fermi arcs in the N\'eel and spiral regimes~\cite{Eberlein2016,Chatterjee2017,Verret2017}, may also survive \cite{Scheurer2018, Bonetti2022gauge, Vasiliou2023}. The mean-field critical temperature $T^*$ then becomes a \textit{crossover} temperature, below which the electronic spectral function develops gaps for momenta in the antinodal region and Fermi arcs near the nodal region.
\begin{figure}[t!]
    \centering
    \includegraphics[width=0.45\textwidth]{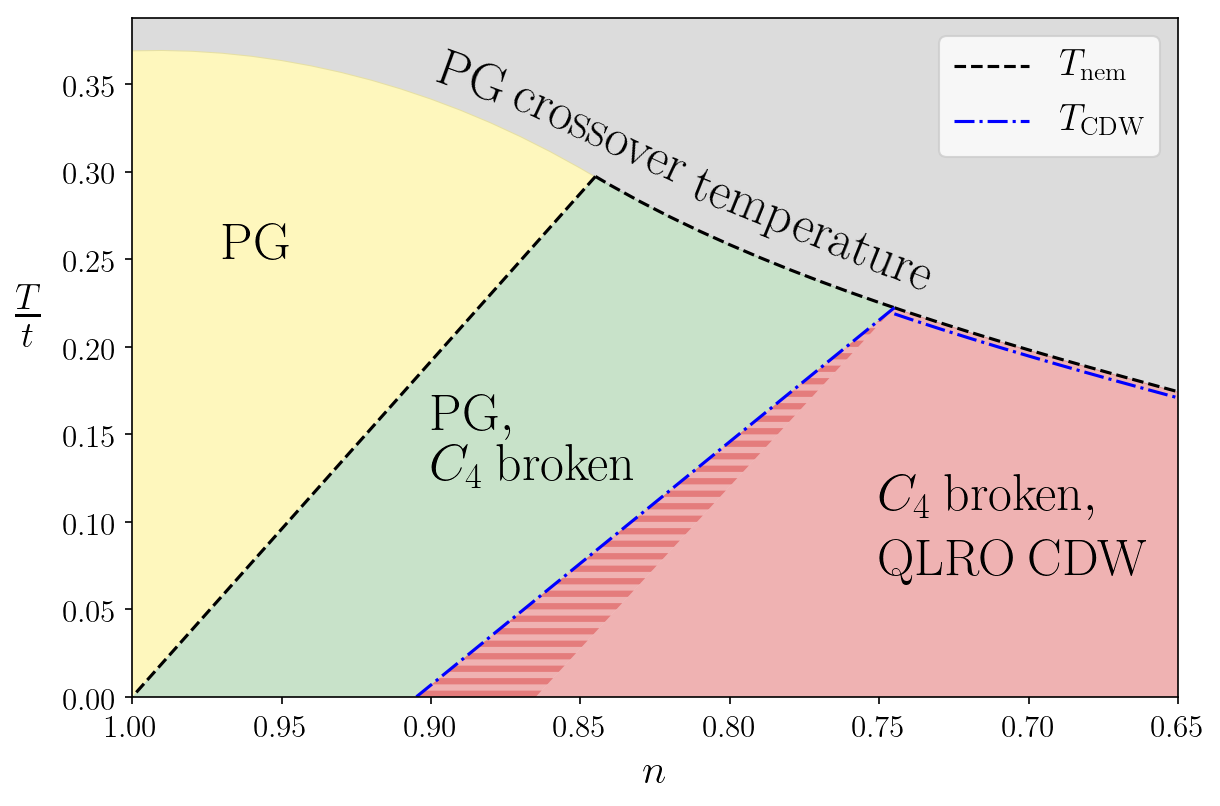}
    \caption{Schematic phase diagram in presence of magnetic order parameter fluctuations in the most relevant density regime, excluding very large hole doping (cf.\  Fig.~\ref{fig: T* phase diag} for the corresponding mean-field phase diagram). Various pseudogap phases with and without nematic ($C_4$ broken) and/or charge density wave (CDW) order are obtained. The hatched red region derives from the multi-spiral phase in mean-field theory, which has the same symmetries as the stripe regime in the presence of thermal fluctuations.}
    \label{fig: Phase diag w fl}
\end{figure}
The magnetic phases obtained in mean-field theory can therefore be interpreted as \textit{pseudogap} (PG) phases at $T > 0$ (see Fig.~\ref{fig: Phase diag w fl}). Whether the ground state remains magnetically ordered depends on the strength of the quantum fluctuations.

The breaking of the discrete (not continuous) $C_4$ rotational symmetry in the spiral, multi-spiral, and stripe phases can survive fluctuations, leaving a \textit{nematic} phase, whose (mean-field) transition temperature $T_\mathrm{nem}$ is indicated in Fig.~\ref{fig: Phase diag w fl}. At low hole doping nematic order sets in at $T_\mathrm{nem} < T^*$, while at larger doping $T_\mathrm{nem}$ and $T^*$ coincide (with $T^*$ of course not being sharply defined in the presence of fluctuations).
Charge density wave (CDW) order, displayed by the multi-spiral and stripe phases, can also survive the presence of fluctuations. Fig.~\ref{fig: GS phase diag} indicates that the incommensurability $\delta$ is a continuous function of the electron density, at least on the mean-field level. This implies that, except for certain special fillings, CDW order is generically \textit{incommensurate}. Such an order has an emergent U(1) symmetry~\cite{Bak1982} and can exist at finite temperature only in the form of topological order in a Berezinskii-Kosterlitz-Thouless (BKT) phase. In Fig.~\ref{fig: Phase diag w fl}, we sketched the mean-field CDW transition temperature $T_\mathrm{CDW}$. The BKT transition temperature for CDW quasi long-range order (QLRO), $T_\mathrm{CDW}^\mathrm{BKT}$, is expected to be lower than $T_\mathrm{CDW}$. The temperature range $T_\mathrm{CDW}^\mathrm{BKT} \leq T \leq T_\mathrm{CDW}$ is a regime of fluctuating short range CDW order. It is possible that at lower temperature CDW fluctuations will lock the period of the charge modulation to a commensurate value, realizing a phase with long range order via an incommensurate to commensurate transition~\cite{Bak1982}.
Spin fluctuations will render the multi-spiral and stripe phase qualitatively identical at finite temperature, as they both break the same symmetries.

%%%%%%%%%%%%%%%%%%%%%%%%%%%%%%%%%%%%%%%%%%%%%%%%%%%%%%%%%%%%%%%%%%%%%%%%%%%%%%%%%%

\section*{Acknowledgements}

We are grateful to A.~Chubukov, A.~Georges, M.~Randeria, and S.~Sachdev for valuable discussions.

P.~M.~B. acknowledges support by the German National Academy
of Sciences Leopoldina through Grant No.~LPDS 2023-06.

%%%%%%%%%%%%%%%%%%%%%%%%%%%%%%%%%%%%%%%%%%%%%%%%%%%%%%%%%%%%%%%%%%%%%%%%%%%%

\begin{appendix}

%%%%%%%%%%%%%%%%%%%%%%%%%%%%%%%%%%%%%%%%%%%%%%%%%%%%%%%%%%%%%%%%%%%%%%%%%%%%

\section{Hamiltonian matrix for {\em P} = 3 and 6}
\label{app: explicit Hamiltonian}
The explicit form of the Hamiltonian~\eqref{eq: H_MF components} for $P=3$ and $P=6$ reads
\begin{equation}
 \mathcal{H}^{(6)\,\mathrm{or}\,(3)}_{\bk,\sigma} =
 \left(
 \begin{array}{llllll}
        \epsilon_{\bk,0} & \Delta^+_{\sigma,1} & \Delta_2^0 & \Delta^+_{\sigma,3}
        & \Delta_4^0  & \Delta^+_{\sigma,5} \\
        \Delta^-_{\sigma,5} & \epsilon_{\bk,1} & \Delta^-_{\sigma,1} & \Delta_2^0
        & \Delta^-_{\sigma,3} & \Delta_4^0 \\
        \Delta_4^0 & \Delta^+_{\sigma,5} & \epsilon_{\bk,2} & \Delta^+_{\sigma,1}
        & \Delta_2^0 & \Delta^+_{\sigma,3} \\
        \Delta^-_{\sigma,3} & \Delta_4^0 & \Delta^-_{\sigma,5} & \epsilon_{\bk,3}
        & \Delta^-_{\sigma,1} & \Delta_2^0 \\
        \Delta_2^0 & \Delta^+_{\sigma,3} & \Delta_4^0 & \Delta^+_{\sigma,5}
        & \epsilon_{\bk,4} & \Delta^+_{\sigma,1} \\
        \Delta^-_{\sigma,1} & \Delta_2^0 & \Delta^-_{\sigma,3} & \Delta_4^0
        & \Delta^-_{\sigma,5} & \epsilon_{\bk,5}
 \end{array}
 \right) \, ,
\end{equation}
where $\epsilon_{\bk,\ell}=\epsilon_{\bk+\ell\bQ_6}$ if $P=6$, and $\epsilon_{\bk,\ell}=\epsilon_{\bk+\ell\bQ_3}$ if $P=3$.

%%%%%%%%%%%%%%%%%%%%%%%%%%%%%%%%%%%%%%%%%%%%%%%%%%%%%%%%%%%%%%%%%%%%%%%%%%%%%%

\section{Landau coefficients}
\label{app: GLT coefficients}
In this section, we report the microscopic expressions for the coefficients of the effective potential in Eq.~\eqref{eq: effective charge spin potential}, as obtained by Taylor expanding the effective action~\eqref{eq: EffectiveBosonicAction}.

The coefficients of the quadratic terms are given by
\begin{subequations}
\begin{align}
 s &= \frac{2}{U} - 2\Pi_0(Q_x) = \frac{2}{U} - 2\Pi_0(Q_y) \, , \\
 r_1& = \frac{2}{U} + 2\Pi_0(2Q_x) = \frac{2}{U} + 2\Pi_0(2Q_y) \, , \\
 r_2 &= \frac{2}{U} + 2\Pi_0(Q_x+Q_y) = \frac{2}{U} + 2\Pi_0(Q_x-Q_y) \, ,
\end{align}
\end{subequations}
where the bare bubble $\Pi_0(q)$ has been defined in Eq.~\eqref{eq: bare bubble}, and $Q_\alpha = (\bQ_\alpha,0)$ for $\alpha = x,y$.

The third order coefficients $b_1$ and $b_2$ are given by
\begin{subequations}
\begin{align}
 b_1 &= 2\int_k G_0(k)\,G_0(k+Q_x)\,G_0(k+2Q_x) \nonumber \\
 &= \{x \leftrightarrow y\} \, , \\
 b_2 &= 4\int_k G_0(k)\,G_0(k+Q_x)\,G_0(k+Q_x+Q_y) \nonumber \\
 &= \{x \leftrightarrow y\} \, .
\end{align}
\end{subequations}
where $\int_k$ indicates an integration over the lattice momentum and a sum over the fermionic Mastsubara frequencies.

The fourth-order coefficients can be conveniently expressed in terms of the integrals
\begin{equation}
\begin{split}
 E_1 & = \int_k G_0(k)\, G_0(k + Q_x)\, G_0(k + 2Q_x)\, G_0(k+Q_x) \, , \\
 E_2 & = \int_k G_0(k)\, G_0(k + Q_x)\, G_0(k)\, G_0(k+Q_x) \, , \\
 E_3 &= \int_k G_0(k)\, G_0(k + Q_x)\, G_0(k)\, G_0(k+Q_y) \, , \\
 E_4 &= \int_k G_0(k)\, G_0(k + Q_x)\, G_0(k + Q_x + Q_y)\, G_0(k+Q_y) \, ,
\end{split}
\end{equation}
in the form
\begin{equation}\label{FinalResult}
\begin{split}
 u_0 &= E_2 + 2E_3 - E_4 \, , \\
 u_1 &= E_2 - 2E_3 + E_4 \, , \\
 u_2 &= 2E_1 - E_2 \, , \\
 u_3 &= 4E_4 \,.
\end{split}
\end{equation}

\end{appendix}

%%%%%%%%%%%%%%%%%%%%%%%%%%%%%%%%%%%%%%%%%%%%%%%%%%%%%%%%%%%%%%%%%%%%%%%%%%%%%%%

\bibliography{main.bib}

\end{document}